\documentclass[%
 aip,
prapplied,
 amsmath,amssymb,
 reprint,
]{revtex4-2}

\usepackage{graphicx}
\usepackage{dcolumn}
\usepackage{bm}
\usepackage[utf8]{inputenc}
\usepackage[T1]{fontenc}
\usepackage{mathptmx}
\usepackage{mathtools}
\usepackage{amsmath}
\usepackage{etoolbox}
\usepackage{color}
\usepackage{siunitx}
\usepackage[normalem]{ulem}
\DeclareSIUnit{\amagat}{amg}
\DeclareSIUnit{\torr}{Torr}

\usepackage[hidelinks]{hyperref}

\hypersetup{
  colorlinks   = true, 
  urlcolor     = blue, 
  linkcolor    = blue, 
  citecolor   = blue 
}

\makeatletter
\def\@email#1#2{%
 \endgroup
 \patchcmd{\titleblock@produce}
  {\frontmatter@RRAPformat}
  {\frontmatter@RRAPformat{\produce@RRAP{\phantom{*}\!\!#1\href{mailto:#2}{#2}
  }}\frontmatter@RRAPformat}
  {}{}
}%
\makeatother

\newcommand{\myaffiliation}{\affiliation}

\newcommand{\ICFO}
{\myaffiliation{ICFO -- Institut de Ci\`encies Fot\`oniques, The Barcelona Institute of Science and Technology, 08860 Castelldefels (Barcelona), Spain}}

\newcommand{\ICREA}{\myaffiliation{ICREA -- Instituci\'{o} Catalana de Recerca i Estudis Avan{\c{c}}ats, 08010 Barcelona, Spain}}

\newcommand{\CSEM}{\myaffiliation{Centre Suisse d'\'Electronique et de Microtechnique (CSEM) SA, Rue Jaquet-Droz 1, 2002 Neuch\^{a}tel, Switzerland}}

\newcommand{\MEGIN}{\myaffiliation{Megin Oy, Keilasatama 5, 02150 Espoo, Finland}}
\newcommand{\VTT}{\myaffiliation{VTT Technical Research Centre of Finland, Tekniikantie 21, 02150 Espoo, Finland}}

\begin{document}

\preprint{AIP/123-QED}

\title{Functionalized mm-scale vapor cells for alkali-metal spectroscopy and magnetometry}

\author{Harini Raghavan}
\thanks{Equal contribution}
\ICFO

\author{Michael C.D. Tayler$^\ast$}
\thanks{Equal contribution}
\ICFO
\email[$\ast$ Electronic mail: ]{michael.tayler@icfo.eu}


\author{Kostas Mouloudakis$^\dagger$} 
\ICFO
\email[$\dagger$ Electronic mail: ]{kostas.mouloudakis@icfo.eu}

\author{Rachel Rae}
\ICFO

\author{Sami L\"ahteenm\"aki}
\VTT
\author{Rasmus Zetter}
\author{Petteri Laine}
\MEGIN

\author{Jacques Haesler}
\author{Laurent Balet}
\author{Thomas Overstolz}
\author{Sylvain Karlen}
\CSEM

\author{Morgan W. Mitchell$^\ddagger$}
\ICFO
\ICREA
\email[$\ddagger$ Electronic mail: ]{morgan.mitchell@icfo.eu}

\date{\today}

\begin{abstract}

\textbf{Abstract:} 
We describe micro-fabricated rubidium vapor cells with integrated temperature-control functionality and demonstrate their suitability for use in miniaturized ultra-sensitive magnetometers. These functionalized vapor cells (FVCs) embody a dual-chamber design in low-conductivity silicon with anti-permeation coatings and    micro-structured thin-film platinum surface traces as resistive heaters and temperature sensors.  
Thermal tests show our ability to control alkali metal distribution within the FVCs, ensuring a clean sensing chamber for optical measurements. Optical absorption spectroscopy is used to correlate the temperature readings with vapor density and to measure buffer gas pressure, of interest for optimizing sensitivity. 
Finally, we demonstrate zero-field resonance magnetometry with \SI{18}{\femto\tesla\per\sqrt\hertz} sensitivity in the \SI{10}{\hertz} to \SI{100}{\hertz} band, limited by laser noise and magnetic shield noise, which indicates that the functionalization does not introduce significant magnetic noise.

\end{abstract}
\maketitle

\section{Introduction}

Optically polarized alkali-metal vapors provide a precise route to measure time, magnetic fields and rotations \cite{Lombardi2007NIST,Tierney2019OPMreview,RevModPhys.89.035002}. With appropriate microfabrication\cite{KnappeShahKitching2,liew2004microfabricated,KnappeShahKitching}, optics and electronics, alkali vapor technologies can be implemented as miniaturized atomic devices\cite{kitching2011atomic, kitchingReview} including chip-scale atomic clocks\cite{Haesler2017}, wearable magnetometer arrays\cite{Boto2018} and portable inertial navigation systems\cite{MacFarlane2003RSTA361}. 
Other applications, including magnetometers for audio-band communications\cite{Lipka2024multiparameter} and nuclear magnetic resonance\cite{Ledbetter2008PNAS,Yu2009CMRA,Blanchard2020JMR,Mouloudakis2023JPCL} are also being studied. 
Large scale deployment of these technologies will require parts that simultaneously satisfy the performance requirements of the application, and are reliably manufacturable in large numbers.

One critical device component is the container for the alkali vapor, commonly termed the ``vapor cell.''\cite{Maurice2022}
A vapor cell must, at a minimum, hermetically isolate the alkali metal and buffer gas from the surroundings, while providing optical transparency. 
The cell geometry contributes to determining important spectroscopic parameters, such as optical path length\cite{PhysRevLett.110.160802,Maurice2022}, beam size and orientation\cite{Yu2023} and the rate of alkali atom-wall collisions. Cell wall thickness also sets the minimum standoff distance to a magnetic field source, an important parameter in magnetometry\cite{RevModPhys.93.035006,Dyer2022}. In many applications, the cell must also be heated, in some cases to $\sim \SI{150}{\celsius}$, to achieve a specific vapor density \cite{gallinet2019atomic,Knappe220921,Lutwak4319292,Affolderbach7140794,Petremand2012,dyer2023real, Martinez2023}. Heater power can be a limiting factor in low-power\cite{PhysRevApplied.18.054039} or low-thermal-load applications.

To ease these many requirements, ``functionalized vapor cells'' (FVCs) have been developed. These incorporate features such as reflective window coatings\cite{Straessle}, integrated heaters\cite{IJP2023, Zipfel2024arxiv, Zhou2024Measurement} and thermistors\cite{Overstolz2014}, lifetime-improving coatings\cite{karlen2017lifetime,Woetzel2013SurfaceCoatTech} and alkali metal preferential condensation zones\cite{Karlen2018}.  FVCs, in particular those implemented using MEMS technology, can enable a high level of integration and wafer-scale manufacturing. To date, however, these technologies have only been used sparsely in applications that require low noise, because noise generated by alkali-metal condensation, cell-body conductivity \cite{GriffithOE2010} and other functional elements of the cell can be significant \cite{Sebbag2021, Ma2022, Jiang2023APL}.

In this paper, we demonstrate a low-noise FVC suitable for zero-field-resonance magnetometry.  The functionalization comprises: (1) a two-chamber high-Z silicon body structure with interconnecting micro-channels for selectively filling the sensing volume with alkali metal; (2) coated internal surfaces to restrict alkali metal permeation; (3) multiple resistive platinum traces on the beam-incident and outgoing cell windows, to apply localized heating as well as to measure the temperature of each cell chamber.  

This FVC builds upon a previous design without platinum-trace functionalization for which we reported a magnetic sensitivity of \SI{20}{\femto\tesla\per\sqrt\hertz} (in the 20-100 \si{\hertz} band) when operated as a single-beam vector magnetometer\cite{Tayler2022APL}.  Based on calculations of others \cite{GriffithOE2010, Sebbag2021, Ma2022, Jiang2023APL} as well as our own\cite{Iivanainen2021JAP}, in this design the thermal magnetic noise from Pt surface elements, Rb droplets, the Si cell body and other sources is expected to be negligible. In practice, the platinum-heater FVC achieves a sensitivity of \SI{18}{\femto\tesla\per\sqrt\hertz} in the same band, all other factors being kept equal. This establishes that the noise due to the on-cell functionalization is small, and that the magnetic sensitivity of our system is competitive with magnetometers in commercial production that employ non-functionalized vapor cells.  The sensitivity appears to be dominated by laser intensity fluctuations.

\begin{figure}[t]
	\centering
\includegraphics[width=\columnwidth]{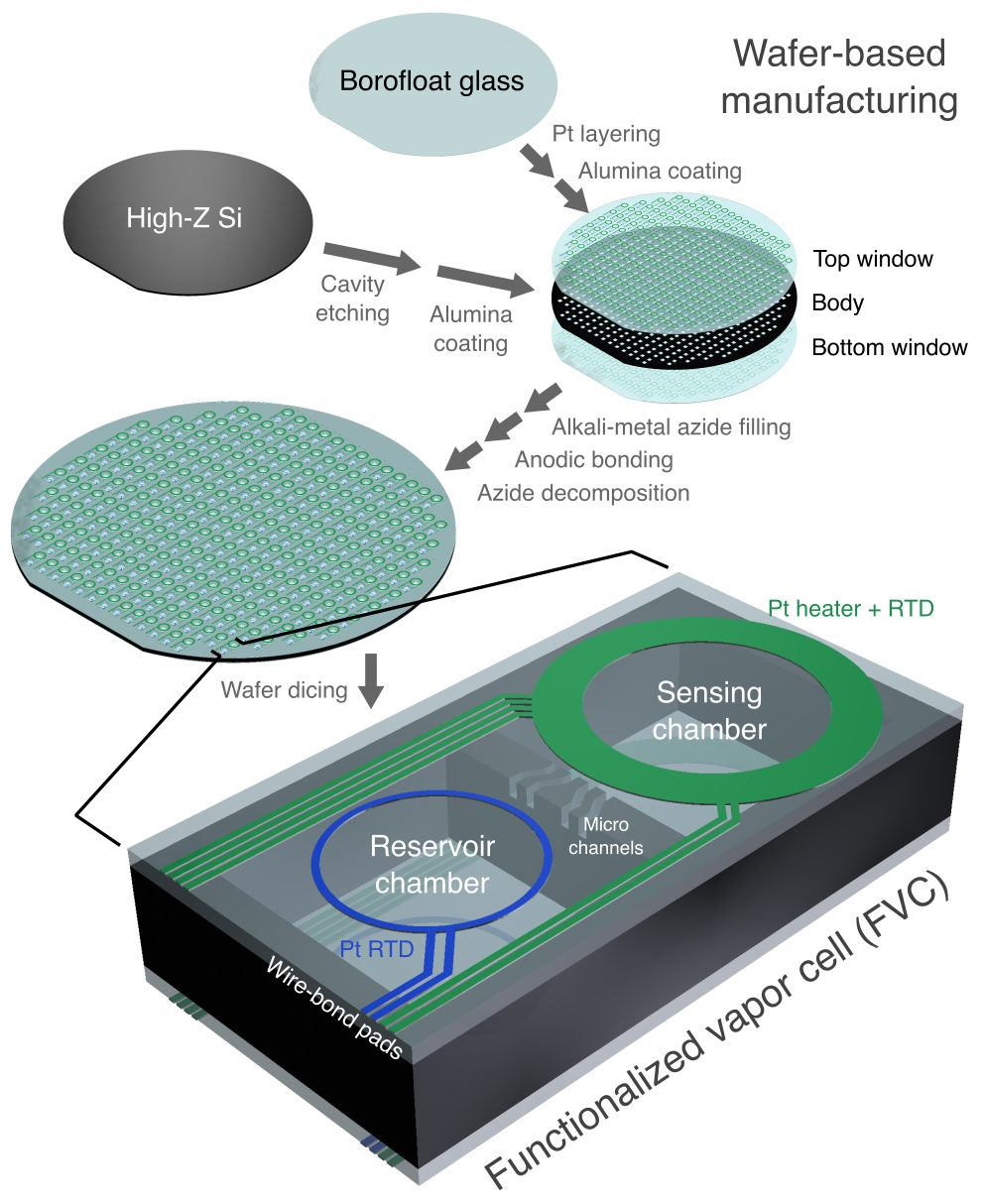}
\caption{Schematic view of a functionalized vapor cell (FVC) for minature alkali-metal sensors, indicating some of the key design elements and stages of the manufacturing process. RTD = resistive thermal detector. 
}\label{fig:dual chamber vapor cell}
\end{figure}

\section{Methods}

\subsection{Functional-vapor-cell (FVC) design}

As illustrated in \autoref{fig:dual chamber vapor cell}, the FVC comprises a hollowed silicon body sandwiched between two glass windows\cite{Yintao_Ma_2022}.  In this design, there are two hollows per cell -- one of these is designated as the ``sensing chamber'', a clean volume in which atoms interact with light, while the other is the ``reservoir chamber'', used for alkali metal storage and other operations, including filling and activation of the cell.  These chambers are connected by an array of micro-channels to allow cross-passage of alkali atoms in the vapor phase.

The exterior window surfaces of the FVC are each patterned with a thin film of platinum metal, as shown in \autoref{fig:dual chamber vapor cell}.  Each pattern contains two sets of traces centered on the sensing chamber as ohmic heaters, plus two traces as resistive thermal detectors (RTDs, see \autoref{sec:RTDcalibration}), one on each chamber for temperature measurement.  All traces are broken out via low-resistance feed wires to pads at the edge furthest away from the sensing chamber.  The traces have a width of \SI{20}{\micro\meter}, while the feed wires are at least \SI{150}{\micro\meter} wide, and the approximate thickness of the Pt film is 300 nm.  

\subsection{Vapor cell fabrication}
The vapor cells are fabricated using standard micro electo-mechanical systems (MEMS) fabrication techniques available at CSEM\cite{Karlen2018}. A total of 200 to 300 cells are produced at wafer-level in a single run. A 1.5-mm-thick, 6-inch-diameter undoped silicon wafer having an electric sheet resistivity of 8-10 \si{\kilo \ohm\centi\meter} (chosen in order to reduce the magnetic signature) is etched by deep reactive ion etching (DRIE). In each cell, voids are created for a sensing and a reservoir chamber (of internal dimensions \SI{4}{\milli\meter} length, \SI{4}{\milli\meter} width each) interconnected by microchannels (0.1 mm width).

After etching, the silicon wafer is coated with a $\approx \SI{20}{\nano\meter}$ layer of Al\textsubscript{2}O\textsubscript{3}\cite{karlen2017lifetime} together with the top and bottom metallized (see \autoref{sec:pt_met}) glass window wafers (Borofloat\textsuperscript{\tiny\textregistered}33 borosilicate glass, \SI{200}{\micro\meter}). The silicon preform is then anodically bonded to the bottom window to create a collection of blind cavities. Droplets of a rubidium-87 azide solution (RbN\textsubscript{3} in H\textsubscript{2}O, 99\% \textsuperscript{87}Rb isotopic abundance) are then dispensed into each of the reservoir cavities by micro-dispensing with a precise pre-determined volume. After evaporation of the liquid component, a thin film of dry \textsuperscript{87}RbN\textsubscript{3} (\SI{2}{\micro\mole}) is left on the window interior surface.  At this point the Al\textsubscript{2}O\textsubscript{3}-coated, metallized top glass window is anodically bonded to the wafer under N\textsubscript{2} atmosphere ($\sim$\SI{3}{\bar}) to seal the cells shut.  The azide salt is then decomposed under UV light into free \textsuperscript{87}Rb metal and N\textsubscript{2} buffer gas.  
Raman spectroscopy was then performed by focusing a probe laser on the reservoir window covered by RbN\textsubscript{3} deposits.  Complete (above 99\%) disappearance of the N\textsubscript{3} Raman stretching vibration signal around \SI{1300}{\per\centi\meter} indicated the complete decomposition of RbN\textsubscript{3} into Rb metal and N\textsubscript{2} gas.  The total N\textsubscript{2} content of the cell is thus given by 1.5 times the number of moles of RbN\textsubscript{3} plus the moles of N\textsubscript{2} present in the cell during the second anodic bonding.  Finally, the wafer is diced to yield individual cells.

Complete decomposition of the azide yields around \SI{160}{\micro\gram} of the free metal inside the cell.  Based on the estimated rate of Rb permeation through the alumina-coated windows, this quantity should suffice to operate the cell for more than 3 years at \SI{200}{\celsius}. 

\subsection{Pt metallization pattern}
\label{sec:pt_met}
Before anodically bonding the windows to the central Si layer, the microstructured Pt layer is deposited on the glass-window outer surface by lift-off over a Ta adhesion layer.  Each window contains heater traces and two RTD traces, one centered on the sensing chamber, the other on the reservoir chamber.  The heater pattern is centered on the sensing chamber to ensure alkali metal condensation occurs preferentially in the reservoir.  This avoids droplets forming in the sensing chamber, which would impede overall light transmission as well as contribute additional thermal magnetic noise\cite{Karlen8409005}.  The four RTDs allow the temperature of each window face of each chamber to be monitored.  All out/return trace wires are placed adjacent to one another to minimize stray magnetic fields due to electrical current.

After anodic bonding assembly but before dicing, electrical resistances ($R$) for each Pt trace on each FVC are tested at room temperature on a flying-lead probe station (Karl Suss, model PA200) in conjunction with a digital multimeter.  For a sample of 132 cells, corresponding to around half of the wafer, more than 100 cells have all trace resistances within 5\% of the respective mean values.  If this figure is taken as the quality threshold, the overall yield of the metallization layers is above 75\%.  However, most of the cells with out-of-spec resistances, defined as $\Delta R/R>1.05$ or $<0.95$, are located together on one side of the wafer, and are the result of over-metallization or ``retention'' after liftoff, which causes shorting between traces.  Ignoring the latter region, the overall FVC yield is above 90\%. Improvements in the liftoff process may therefore increase the yield of FVCs substantially.

\subsection{Electrical connections}
\label{sec:pcb}

After dicing, individual cells are held within a ``carrier'' printed circuit board (PCB) made of standard FR4 fiberglass (PCBway, Hangzhou, China).  Electroless nickel-immersion-gold (ENIG-)plated copper pads of \SI{1}{\micro\meter} thickness are placed at the edge of the PCB, spaced with a pitch of 0.5 mm to provide electrical feed wires to connect to the RTD and heater traces.
The ENIG PCB and Pt cell breakout pads are connected to one another by aluminum wedge bonds.  Aluminum bond wire (diameter \SI{25}{\micro\meter}, AMETEK ALLOY: 1\%Si/Al) is placed manually at room temperature using a bench-top apparatus (West-Bond model 7476e-79) as follows.  First, both breakout surfaces are cleaned with solvent and then blown dry with compressed air.  For each connection, the initial bond is made to the ENIG breakout pad (0.4\,W ultrasonic power; 40\,ms time; force 21\,g) and the second bond to the platinized surface (0.4\,W ultrasonic power; 40\,ms time; force 25\,g).  

Wire bonding is not the only bonding technique applicable to thin-layer Pt.  Other methods commonplace in large-scale MEMS manufacturing could be used, e.g., flip-chip bonding.

\subsection{RTD characterization}
\label{sec:RTDcalibration}
The electrical resistance of each RTD depends linearly on its temperature ($T$) according to:
\cite{Mesures1969} 
\begin{equation}
 R(T) = R_0 [1 +\alpha (T - T_0)],
 \label{eq:ResistancevsT}
\end{equation}
where $R_0$ is the known value of resistance at a fixed reference temperature $T_0$ and $\alpha$ is the temperature coefficient of the material. 

The value of $R_0$ can in principle be calculated from the material dimensions using the formula $R_0= \rho L/A$, where $L$ is the length, $A$ is the cross-sectional area of the RTD trace, and $\rho$ is the resistivity of the metal.  However, because neither $L$ nor $A$ are known to great accuracy, an experimental calibration of $\alpha$ and $R_0$ is performed.  

To calibrate the RTDs, the MEMS cell and breakout-PCB assembly is submerged in a bath of mineral oil (of volume $\sim$\SI{100}{\centi\meter\cubed}) on a hot plate, which heats the oil to a steady temperature of around \SI{140}{\celsius}.  The hot plate is then switched off and the oil temperature and resistance of each RTD is continuously measured during cool-down, which takes approximately 30 minutes, all the way to room temperature.  The temperature was measured using a calibrated K-type thermocouple also immersed in the oil bath.  A value for $T_0$ can be freely chosen and $R_0$ and $\alpha$ are then found by linear regression.
Using room temperature as a reference ($T_0 =$ 298 K) we find 
$R_{0,\rm sens.} =$ \SI{990}{\ohm} and $\alpha_{\rm sens.} =$ \SI{0.274}{\per\kelvin} for the sensing chamber and $R_{0,\rm resv.} =$ \SI{230}{\ohm} and $\alpha_{\rm resv.} =$ \SI{0.269}{\per\kelvin} for the reservoir.  The values of $\alpha$ agree closely with the literature value of \SI{0.272}{\per\kelvin} for thin-film Pt. The RTD calibration is consistent across multiple vapor cells from the same wafer batch.

\subsection{Absorption spectroscopy}
\label{sec:spectroscopymethods}
Optical transmission through the sensing chamber is measured in an unshielded (Earth's field) environment. The light source is a distributed Bragg reflector (DBR) laser emitting a continuous beam near the D\textsubscript{1} transition line of \textsuperscript{87}Rb ($\sim$\SI{795}{\nano\meter}). A photodetector (Thorlabs DET36A2) is located on the other side of the cell to detect the transmitted power.

The dependence of the transmission on wavelength is measured by slowly scanning the laser temperature so as to scan the light frequency across the D\textsubscript{1} line (\SI{120}{\giga\hertz} in \SI{5}{\second}). The linearity between temperature and frequency is verified using a calibrated high-resolution wavelength meter (HP 86120B), and the range is calibrated using the Doppler-broadened D\textsubscript{1} line of a natural-abundance-Rb spectroscopy reference cell (Thorlabs GC25075-RB) at room temperature.  The DBR output couples into an optical fiber and is then conditioned with a linear polarizer to produce a pure linear polarization. 

This linearly polarized beam is equally split among two beams. One passes through the FVC, and the other through free space; each is detected at a separate photodiode. The ratio of the two intensities provides an absolute measure of light transmission through the FVC, and is insensitive to changes in beam power.

The shape of the absorption peak in the spectrum is fit to a simple model that describes collisions between the alkali-metal atoms and the  N\textsubscript{2} buffer gas in the FVC.  The model assumes a homogeneously broadened medium, the absorption cross-section is approximated with a Lorentzian function\cite{2011PhDT........41V}, and the absorption is described by the Beer-Lambert law.  The resulting transmission is 
\begin{equation}
\mathbb{T} (\nu) \equiv \frac{P_\mathrm{out}}{P_\mathrm{in}} = \exp \left[-n_\mathrm{Rb} r_ecf_{\rm{osc}} L \frac{\Delta \nu/2}{(\nu-\nu_0)^2+(\Delta \nu/2)^2}\right] ,
\label{eq:Lorentzian function}
\end{equation}
where $P_\mathrm{in}$ is the input power, equal to the reference beam power, $P_\mathrm{out}$ is the output power, $n_\mathrm{Rb}$ is the alkali number density, $r_e \approx \SI{2.8 e-15}{\meter}$ is classical radius of the electron, $c$ is the speed of light, $L$ is the internal path length through the cell, $f_\mathrm{osc}\approx 0.332$ is the oscillator strength for the D\textsubscript{1} line,  $\Delta \nu$ is the FWHM pressure-broadened optical linewidth, and $\nu-\nu_0$ is the detuning of the laser light (of frequency $\nu$) from the pressure-shifted optical frequency $\nu_0$ associated with the $\rm{D}_1$ transition from the ground to the excited electronic state.

\subsection{Zero-field-resonance (ZFR) magnetometry}
\label{sec:ZFRmagnetometry}

Performance of the FVC for single-beam vector magnetometry \cite{shah2007subpicotesla} is tested on the laboratory bench setup illustrated in \autoref{fig:magnetometry}a.  The FVC is mounted on a FR4 PCB structure containing traces for the heater/RTD connections and wire-bond pads, similar to that described in \autoref{sec:pcb} but smaller in size.  Above (the top window) and below (the bottom window of) the FVC is placed a layer of insulation (Airloy HT flexible foam) and a set of miniature biplanar printed-circuit coils for local magnetic field control along three orthogonal axes, with the origin centered upon the sensor chamber. The coils' construction and characterization are described in detail by Tayler et al. \cite{Tayler2022APL}  
These components contain holes for beam entry and exit, and are held together in a machined polyetheretherketone (PEEK) casing of outer dimensions \SI{25}{\milli\meter} by \SI{20}{\milli\meter} by \SI{40}{\milli\meter}.  In this environment the FVC heater is driven to a sensing-chamber RTD temperature above \SI{150}{\celsius} using approximately \SI{1}{\watt} continuous AC power at a frequency of \SI{120}{\kilo\hertz}.  No active temperature stabilization is used.

The PEEK-encased module is tested in a multilayer magnetic shield (Twinleaf MS-1LF). The shield passively screens out the Earth's magnetic field in the interior, and shim coils integrated within the innermost layer null the residual magnetic field to below \SI{1}{\nano\tesla} when fed by a low-noise current source (Twinleaf CSB-10).  According to the manufacturer specifications, the magnetic noise of the shield at low frequencies (below \SI{100} {\hertz}) is expected to be about \SI{1} {\femto\tesla\per\sqrt\hertz}. A commercial atomic magnetometer (QuSpin Inc.\ QZFM-gen2 \cite{quspinQZFMGen3}) verifies that the noise level is below \SI{7} {\femto\tesla\per\sqrt\hertz}.

All optical components are located outside of the magnetic shield, namely the light source, polarizers and photodetectors.  The source used is a VCSEL of $\sim$\SI{1}{\milli\watt} optical power tuned to the center of the pressure-shifted and pressure-broadened \textsuperscript{87}Rb $\rm{D}_1$ line.
A zero-order half-wave plate, a polarizing beam splitter and a zero-order quarter-wave plate are placed in series along the beam path to control the laser power and to circularly polarize the light before reaching the PEEK module.  Light on the other side of the module, i.e., having passed through the FVC, is focused by a convex lens onto the active area of a silicon photodetector (Thorlabs PDA36A2).

The quasi-static zero-field resonance signal, i.e., the transmission as a function of applied transverse field, is obtained by first nulling the three field components as seen by the vapor, then repeatedly sweeping the transverse field from \SI{-800}{\nano\tesla} to \SI{800}{\nano\tesla} in \SI{0.2}{s} while recording the resulting transmitted power.  Here and below, ``transverse'' indicates a direction orthogonal to the laser propagation direction. See \autoref{fig:magnetometry}b for representative signals over the range \SI{-250}{\nano\tesla} to \SI{250}{\nano\tesla}.
 
For magnetic noise measurements, the ZFR magnetometer is operated in an open-loop mode as follows. A constant-amplitude, sine-modulated transverse magnetic field is applied in a near-zero background field; the field is supplied through one of the miniature coils (field amplitude \SI{35}{\nano\tesla}, frequency \SI{790}{\hertz}). 
The photodiode response to the transmitted light is demodulated at the modulation frequency using a lock-in amplifier (LIA, Stanford Research Systems SR830, time constant \SI{1} {\milli \second}, filter order 24 dB/octave, sensitivity setting \SI{500}{\milli\volt}) and the quadrature component is digitized (National Instruments PCI-4462, 24-bit, \SI{200}{ksps}) and stored on a computer.

\newcommand{\SNR}{\zeta}

We measure the magnetic sensitivity, i.e., the single-sided power spectral density (PSD) $S_B(f)$ or amplitude spectral density (ASD) $S_B^{1/2}(f)$ of the sensor noise expressed as an equivalent magnetic field noise at frequency $\nu$, as follows. Together with the modulation field, we apply a sinusoidally-varying test signal $B(t) = B_\mathrm{test} \cos(2 \pi f t)$ in the transverse direction, with amplitude $B_\mathrm{test} \approx \SI{50}{\pico\tesla}$ and frequency $f$ between \SI{5}{\hertz} and \SI{200}{\hertz}. We acquire time traces, of duration $\tau = \SI{1}{\second}$, of the LIA output $V(t)$, as described in the preceding paragraph. We compute the corresponding PSD $S_V(f)$ by discrete Fourier transform with a Hann window, and average the PSD of several traces to reduce statistical uncertainty. From the resulting averaged spectrum we obtain the signal-to-noise ratio $\SNR(f)$, defined as the ratio of the peak to the noise background in $S_V(f)$. The sensitivity is then calculated as 
\begin{equation}
S_B(f) = \frac{B_\mathrm{test}^2}{\SNR(f)} \frac{\tau}{2w_H} ,
\label{eq:mag_sens}
\end{equation}
where $\tau$ is the measurement time and $w_H = 3/2$ is the noise power bandwidth of the Hann window.

\section{Results}

\subsection{Thermal tests}

\begin{figure}
\centering
\includegraphics[width=\columnwidth]{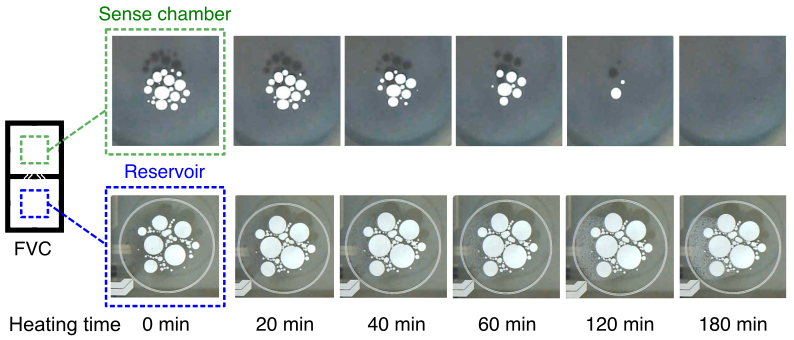}
\caption{Thermal redistribution of metallic \textsuperscript{87}Rb in the MEMS cell.  After UV-decomposition of isotopically enriched \textsuperscript{87}Rb azide, both of the interconnected chambers of the cell contain Rb droplets.   Heating to around \SI{200}{\celsius} using the cell-window-mounted Pt traces induces a thermal gradient (sensing RTD $\sim$\SI{40}{\celsius} hotter than the reservoir RTD) to drive Rb completely into the reservoir chamber, a one-time process which takes around 3 h.  
}
\label{fig:evaporation in sci chamber}
\end{figure}

The heating circuit can be operated consistently to a sense-chamber RTD temperature above \SI{220}{\celsius}, reached after only 10 minutes' heating from a cold start.

To avoid condensation of metal in the sensing chamber, the coldest point on the inner surface of that chamber should be hotter than the coldest point on the inner surface of the reservoir chamber.
For a diced and individually packaged cell, this is consistent with expectations as shown in \autoref{fig:evaporation in sci chamber}.  The initial distribution of metal is determined by the conditions during the azide decomposition step; the condensation occurs (1) at the centers of the windows because these locations are the ones most exposed to heat loss compared to the rest of the cell/wafer, and (2) in both chambers because of cross-chamber atomic migration.  After this, when the Pt trace heaters are first activated to create a temperature differential between the chambers, the metal in the (warmer) sensing chamber migrates completely to the (cooler) reservoir.  This indicates that the coldest points in the cell are the reservoir windows.  We note that the sensing chamber RTD is  closer to the heater traces than is the reservoir chamber RTD, and thus the relevant temperature differential for condensation may be lower than the differential indicated by the RTDs.

Around three hours of heating is needed for complete diffusion through the interconnecting microchannels.  A clean optical window persists thereafter in the sensing chamber when the heater is turned off, at room temperature.

\subsection{Absorption spectroscopy}

\begin{figure}[tp]
\centering
\includegraphics[width=1.\columnwidth]{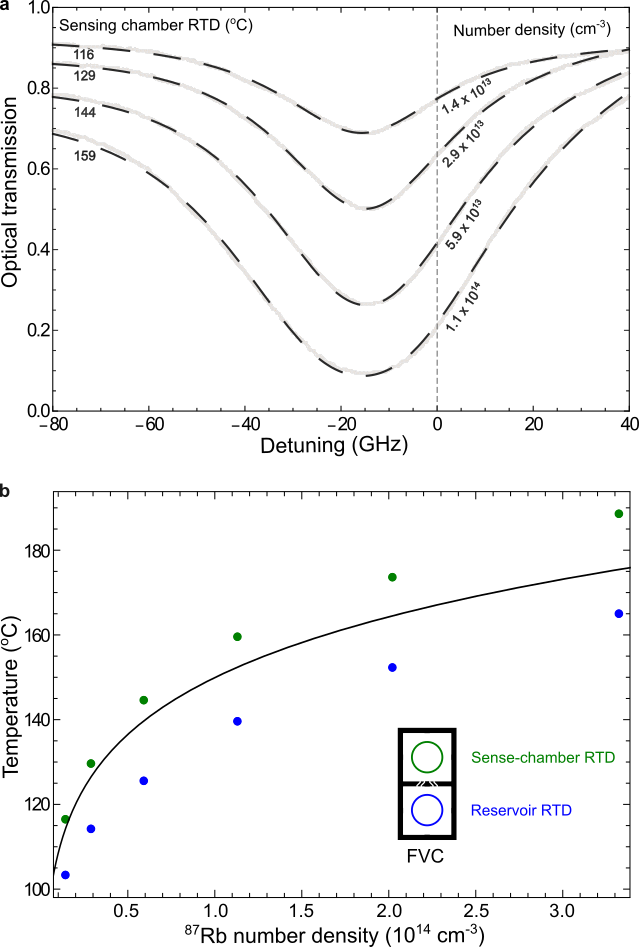}
\caption{Optical transmission through the FVC sensor chamber in the Earth's magnetic field: (a) Temperature and tuning dependence of transmission for a $\sim$\SI{10}{\micro \watt} linearly polarized beam.  Gray solid lines represent experimental data while dashed black lines represent fits using the model of \autoref{eq:Lorentzian function}. The dashed vertical line at zero detuning marks the center of the $\rm{D}_1$ resonance in a spectroscopy reference cell (\SI{794.978}{\nano\meter}); (b) Rb number densities $n_{\rm{Rb}}$ obtained from the fitted D\textsubscript{1} absorption lines of the spectra vs.\ $T$.  The solid curve shows $n_{\rm{Rb}}=\SI{1.506e26}{\per\centi\meter\cubed\kelvin} \times (1/T) \times 10^{-\SI{4040}{\kelvin}/T}$, from\cite{AlcockCMQ1984, CRCHandbook97thEdition}, which agrees with older measurements \cite{KillianPR1926, SeltzerThesis}. See also \cite{SinghPDF2008}. 
} \label{fig:absorption Spectrum} 
\end{figure} 

\begin{figure*}[tp]
\centering
\includegraphics[width=1.\textwidth]{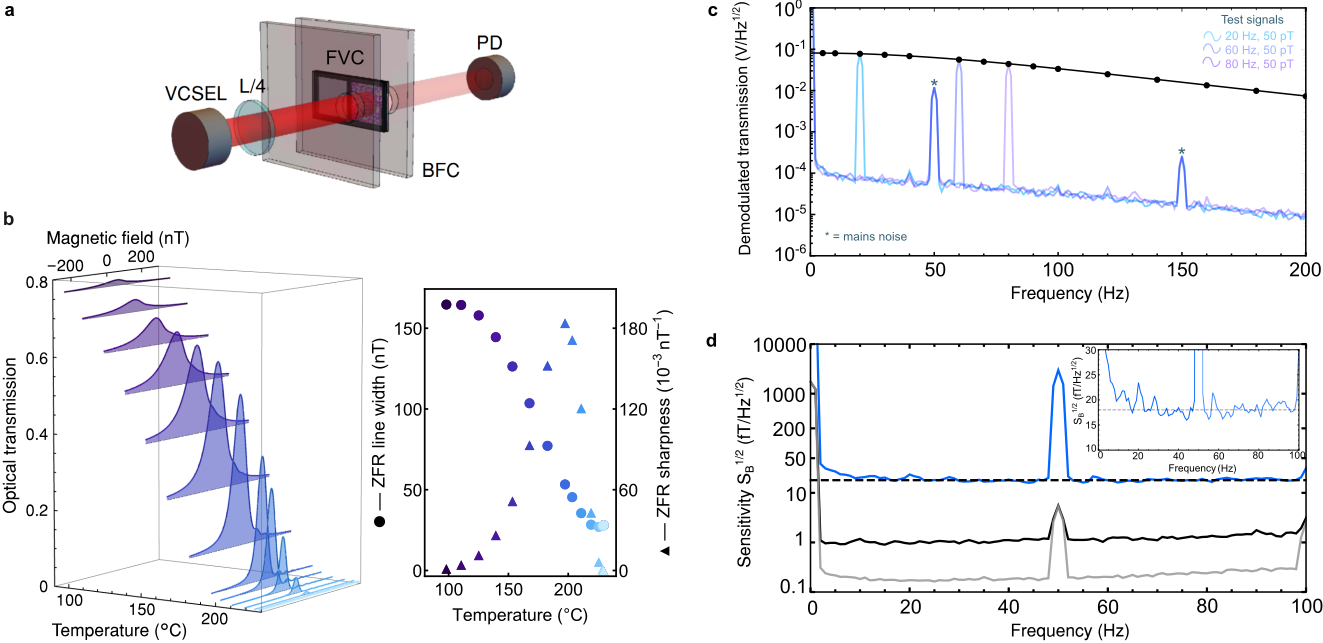}
\caption{Zero-field-resonance magnetometry with the FVC.  
(a) Schematic of the experimental setup for measuring the transmission of \SI{795} {\nano\meter} light.  Linearly polarized light is emitted from a VCSEL and is circularly polarized using a quarter-wave plate (L/4). The beam is then passed through the FVC sensing chamber and collected at a photodetector (PD). Biplanar field coils (BFC) apply compensation and modulation fields. 
(b)Transmission versus temperature and transverse field, acquired quasi-statically with a \SI{5}{\hertz} linear ramp. (c) Frequency response of the OPM. Black dots indicate peak $\sqrt{S_V(f)}$ recorded for a sinusoidal applied transverse field of amplitude \SI{50}{\pico\tesla}, for different frequencies. The solid black curve shows an interpolation of the response. (d) Amplitude spectral density of the open-loop demodulated quadrature signal $\sqrt{S_B(f)}$ when modulating the transverse magnetic field under conditions for highest sensitivity, see \autoref{sec:ZFRmagnetometry}.  The blue curve shows total noise in field units; the dashed horizontal line indicates $\sqrt{S_B(f)} = \SI{18}{\femto\tesla\per\sqrt\hertz}$; solid black curve shows measured electronic noise with the laser off, accounting for the noise due to the dark current of the photodetector, the electronic noise of the lock-in amplifier and the noise of the data acquisition system; gray curve shows the noise of the data acquisition system.  The spectra shown are averages of $100$ acquisitions, each of \SI{1}{\second} duration. A Hann window is applied. Inset shows the same $\sqrt{S_B(f)}$ spectrum on a linear scale. }
\label{fig:magnetometry}
\end{figure*}

The parameter estimates are made by fitting a Lorentzian line shape profile (\autoref{eq:Lorentzian function}) to the D\textsubscript{1}-wavelength absorption profiles at different temperatures: see \autoref{fig:absorption Spectrum}a for a sample of four representative spectral fits.  The individual fitted spectra provide the pressure-shifted resonance frequency ($\nu_0$)
from the reference Rb D\textsubscript{1} frequency, the pressure-broadened line width (FWHM, $\Delta \nu$), and $n_{\mathrm{Rb}}$.  The width is independent of temperature over the regime studied.  We find the fitted value $\Delta \nu = $ \SI{40.1}{\giga\hertz} corresponds to a N\textsubscript{2} buffer gas density of around \SI{2.25}{\amagat}  when dividing by the literature value\cite{PhysRevA.56.4569} for the broadening rate,\SI{-17.8 \pm 0.3}{GHz/\amagat}.
However, the total N\textsubscript{2} gas density should be \SI{2.62}{\amagat}, as the sum of the backing pressure during anodic bonding (\SI{1.31}{\amagat}) and the gas released from the azide (also \SI{1.31}{\amagat}, assuming 100\% decomposition).
The source of the 14\% discrepancy between the two values is unknown.
The \textsuperscript{87}Rb number density ranges from  
$n_\mathrm{Rb}=1.4\times 10^{13}$ \si{\per\centi\meter\cubed} at the lowest RTD temperature (sensing chamber RTD, \SI{116} {\celsius}) to $n_\mathrm{Rb}=3.3\times 10^{14}$ \si{\per\centi\meter\cubed} at the highest (sensing chamber RTD, \SI{159} {\celsius}).

\subsection{Magnetometry}

The magnetic sensitivity is a function of several experimental parameters, including  cell temperature, buffer gas pressure, laser power, path length, modulation strength, and modulation frequency.  
Global optimization of these parameters requires a multi-dimensional search, because such parameters affect the optical/atomic response in overlapping ways. For example, increasing the buffer gas pressure reduces the magnetic linewidth and reduces the optical depth, whereas increasing temperature reduces the magnetic linewidth and increases the optical depth.
One can get a rough idea of the optimal sensing conditions by considering the ``sharpness,'' defined as the height of the transmission peak divided by its FWHM width.  In a quasistatic model, the sensitivity $S_B^{1/2}$ is inversely proportional to the sharpness.  As seen in \autoref{fig:magnetometry}b, the sharpness first increases with increasing temperature, because the height increases while the linewidth decreases, reaches a maximum at \SI{197}{\celsius} (sensing chamber RTD), and then decreases as both the height and linewidth decrease. This temperature of optimal sharpness is  close to but not the same as the observed optimal temperature for sensitivity (described below), found at RTD temperatures \SI{210}{\celsius} (sensor) and \SI{180}{\celsius} (reservoir). Maintaining this optimal temperature requires less than \SI{1}{\watt} of heater power.

Using the method described in \autoref{sec:ZFRmagnetometry}, we measure the sensor equivalent magnetic noise, i.e., the sensitivity $S_B(f)$. The optimized result, at the temperature given above and with a modulation amplitude close the ZFR linewidth of $\approx \SI{45}{\nano\tesla}$, 
is shown in \autoref{fig:magnetometry}d, and gives a nearly-constant $S_B(f) \approx\SI{18} {\femto\tesla\per\sqrt\hertz}$ in the 20 Hz to 100 Hz band. The electronic noise contribution, from the photodetector, LIA and DAQ, is $\approx \SI{1} {\femto\tesla\per\sqrt\hertz}$.

This sensitivity is slightly better than what we previously reported with non-metallized cells of similar dimensions \cite{Tayler2022APL}. This confirms that functionalization by surface metallization can be applied for high-sensitivity magnetometry.

\newcommand{\LPF}{PF}
As just described, the temperature, and thus $n_\mathrm{Rb}$, that optimizes sensitivity is higher than the temperature that optimizes sharpness. This suggests the sensor noise has a significant contribution from laser power fluctuations (\LPF) at the coil modulation frequency, as has been observed in other ZFR OPMs\cite{Krzyzewski2019}. A steady-state, classical noise model for the \LPF 
 ~contribution to detector noise is
\begin{equation}
S_B^\mathrm{\LPF} = \left( \frac{dV}{dB} \right)^{-2} S_V^\mathrm{\LPF}(f_\mathrm{mod}) = \left( \frac{dV}{dB} \right)^{-2} \mathbb{T}^2 \eta^2 S_P(f_\mathrm{mod}),
\end{equation}
where $V$ is the photodetector signal, ${dV}/{dB}$ is the steady-state responsivity to transverse field, proportional to the sharpness, $S_V^\mathrm{\LPF}(f_\mathrm{mod})$ is the noise contributed by \LPF{} at the modulation frequency $f_\mathrm{mod}$, $\eta$ is the photodetector responsivity and $S_P$ is the \LPF{} PSD. In this model, increasing $n_\mathrm{Rb}$ reduces $S_B^\mathrm{\LPF}$ even at maximum sharpness, because it reduces the transmission $\mathbb{T}$ and thus $S_V^\mathrm{\LPF}$. In contrast, magnetic noise from, e.g., thermal currents in the Pt traces, or electronic noise in the detection, would not have this $\mathbb{T}$ dependence. The sharpness optimum is, however, obtained under quasi-static conditions, whereas the sensitivity is measured with a modulation that is both fast relative to spin relaxation, and comparable in strength to the resonance width. This leaves open other possible explanations for the different optimization temperatures.

\section{Discussion and Conclusion}

A key result is the demonstration that the Pt film contributes negligible magnetic noise when the cell is heated to operating temperatures around \SI{200}{\celsius}.  Because the film occupies a much smaller volume than the cell itself, and also because the Pt-liftoff feature size can be made smaller, around \SI{5}{\micro\meter}, we expect this heating method can scale favorable to even-smaller vapor cells, which could include those carved from glass-only structures by direct laser writing\cite{LuciveroLWVC,zanoni2024laserwritten}.  The small-size feature is relevant for applications that detect fields from localized sources, because in these cases the sensor-to-source standoff distance strongly affects the detected field strength.

Pt resistance thermometers are an accurate and industry-standard technology.  Precise placement of these functional parts allows us to quantify the temperature differential between the two  chambers of the FVC, which controls the alkali-metal distribution.  The combined use of thin-film heaters and RTDs allows us to avoid condensation of alkali metal in the sensor chamber, which is a known and significant issue with traditional azide-filled vapor cells\cite{GriffithOE2010}.

The optical absorption spectroscopy gives direct and cell-specific information about important parameters of the atomic vapor, including: 1) the correspondence between atomic vapor density ($n_{\mathrm{Rb}}$) and RTD temperature; 2) the  N\textsubscript{2} buffer gas density. Knowing these parameters facilitates optimization of the magnetic sensitivity\cite{mouloudakis2024spin}.  
These measurements can be performed for individual cells as described in \autoref{sec:spectroscopymethods} or in principle on-wafer 
\cite{Karlen2017AppliedSpec}, to determine each FVC's viability at an early stage, prior to sensor assembly and packaging.  In-sensor measurements of the optical transmission can also inform about ageing of the cell contents\cite{karlen2017lifetime} and leakage.

\autoref{fig:absorption Spectrum}b compares the spectroscopically-measured rubidium number density $n_\mathrm{Rb}$ against the Killian formula. We note that in equilibrium, it is the coldest point on the interior of the cell that determines the vapor pressure. This will be close to the reservoir RTD temperature, which is near to the location in which the metal in fact accumulates (see \autoref{fig:evaporation in sci chamber}) and well below the temperature of the sensing chamber RTD, which is adjacent to the heater. Across the range measured, the Killian formula agrees with the observed $n_\mathrm{Rb}$ at a temperature about $\SI{10}{\celsius}$ \textit{above} the reservoir RTD reading. There is thus a significant discrepancy between the Killian formula and observation. Similar discrepancies have been reported elsewhere \cite{SeltzerThesis, WuAO1986}  for Rb vapor cells with externally-mounted temperature sensors.

In the implementation described here, the heater, RTD traces and feed wires occupy much of the window surface area, and cannot cross without creating short circuits. Smaller cells, or cells with more elaborate functionalization, might thus benefit from multi-layer functionalization, e.g., by direct writing\cite{Edri2021ACSappliedmaterials}, analogous to multi-layer printed circuit boards that achieve dense and complex wiring patterns.

In summary, we have described and characterized functionalized MEMS vapor cells for alkali-metal-vapor sensing applications, in particular for magnetometry.  The cells are lithographically fabricated on \SI{150}{mm}-diameter wafer materials starting from a DRIE, high-Z-silicon cell body with two interconnecting chambers that allocate ``dirty'' filling and ``clean'' sensing regions for the alkali metal.  On top and bottom faces of this body, we add glass windows outfitted with exterior thin-film-Pt resistive heaters and RTDs, plus other elements, produced with inexpensive industry-standard processing (liftoff, anodic bonding). We describe thermal management of vapor density and Rb metal distribution among the  chambers using the integrated heaters and RTDs, and measurement of cell contents by absorption spectroscopy.
Using the functionalized vapor cell we demonstrate a magnetic sensitivity of \SI{18}{\femto\tesla\per\sqrt\hertz} using a single-beam SERF OPM, thus showing that industry-standard, inexpensive, wafer-scale processes important for deployment can give the required performance for demanding quantum sensing applications.

\section*{Data availability}

Numerical data appearing in \autoref{fig:absorption Spectrum} and \autoref{fig:magnetometry} of the published article are freely available via the OPENAIRE data base, hosted on Zenodo: URL \url{https://zenodo.org/doi/10.5281/zenodo.13125374}.

\section*{Acknowledgments}

The work described is funded by: The EU's European Innovation Council project OPMMEG (101099379);
The Spanish Ministry of Science MCIN with funding from European Union NextGenerationEU (PRTR-C17.I1) and by Generalitat de Catalunya ``Severo Ochoa'' Center of Excellence CEX2019-000910-S; 
the Spanish Ministry of Science projects SAPONARIA (PID2021-123813NB-I00), MARICHAS (PID2021-126059OA-I00), SEE-13-MRI (CPP2022-009771) plus RYC2022-035450-I, funded by MCIN/AEI /10.13039/501100011033; 
Generalitat de Catalunya through the CERCA program;  
Ag\`{e}ncia de Gesti\'{o} d'Ajuts Universitaris i de Recerca Grant Nos. 2017-SGR-1354 and 2021 FI\_B\_01039; 
Fundaci\'{o} Privada Cellex; 
Fundaci\'{o} Mir-Puig; Grant FJC2021-047840-I funded by MCIN/AEI/ 10.13039/501100011033 and by the European Union project ``NextGenerationEU/PRTR.'' 
The project that gave rise to these results received the support of a fellowship from the ”la Caixa” Foundation (ID 100010434). The fellowship code is LCF/BQ/DI24/12070013.
Funded by the European Union (ERC, Field-Seer, 101097313 and RIA project QUANTIFY, 101135931). We acknowledge the European Union’s Horizon 2020 research and innovation programme under project macQsimal (Grant Agreement No.\ 820393).;
Views and opinions expressed are those of the authors and do not necessarily reflect those of the European Union or the European Research Council Executive Agency.
The authors also thank Tapani Makkonen for finite-element-modeling simulations related to the thermal performance of the FVC.

\bibliography{bibliography}

\begin{thebibliography}{62}%
\makeatletter
\providecommand \@ifxundefined [1]{%
 \@ifx{#1\undefined}
}%
\providecommand \@ifnum [1]{%
 \ifnum #1\expandafter \@firstoftwo
 \else \expandafter \@secondoftwo
 \fi
}%
\providecommand \@ifx [1]{%
 \ifx #1\expandafter \@firstoftwo
 \else \expandafter \@secondoftwo
 \fi
}%
\providecommand \natexlab [1]{#1}%
\providecommand \enquote  [1]{``#1''}%
\providecommand \bibnamefont  [1]{#1}%
\providecommand \bibfnamefont [1]{#1}%
\providecommand \citenamefont [1]{#1}%
\providecommand \href@noop [0]{\@secondoftwo}%
\providecommand \href [0]{\begingroup \@sanitize@url \@href}%
\providecommand \@href[1]{\@@startlink{#1}\@@href}%
\providecommand \@@href[1]{\endgroup#1\@@endlink}%
\providecommand \@sanitize@url [0]{\catcode `\\12\catcode `\$12\catcode `\&12\catcode `\#12\catcode `\^12\catcode `\_12\catcode `\%12\relax}%
\providecommand \@@startlink[1]{}%
\providecommand \@@endlink[0]{}%
\providecommand \url  [0]{\begingroup\@sanitize@url \@url }%
\providecommand \@url [1]{\endgroup\@href {#1}{\urlprefix }}%
\providecommand \urlprefix  [0]{URL }%
\providecommand \Eprint [0]{\href }%
\providecommand \doibase [0]{https://doi.org/}%
\providecommand \selectlanguage [0]{\@gobble}%
\providecommand \bibinfo  [0]{\@secondoftwo}%
\providecommand \bibfield  [0]{\@secondoftwo}%
\providecommand \translation [1]{[#1]}%
\providecommand \BibitemOpen [0]{}%
\providecommand \bibitemStop [0]{}%
\providecommand \bibitemNoStop [0]{.\EOS\space}%
\providecommand \EOS [0]{\spacefactor3000\relax}%
\providecommand \BibitemShut  [1]{\csname bibitem#1\endcsname}%
\let\auto@bib@innerbib\@empty
\bibitem [{\citenamefont {Lombardi}, \citenamefont {Heavner},\ and\ \citenamefont {Jefferts}(2007)}]{Lombardi2007NIST}%
  \BibitemOpen
  \bibfield  {author} {\bibinfo {author} {\bibfnamefont {M.~A.}\ \bibnamefont {Lombardi}}, \bibinfo {author} {\bibfnamefont {T.~P.}\ \bibnamefont {Heavner}},\ and\ \bibinfo {author} {\bibfnamefont {S.~R.}\ \bibnamefont {Jefferts}},\ }\bibfield  {title} {\enquote {\bibinfo {title} {{NIST Primary Frequency Standards and the Realization of the SI Second}},}\ }\href {https://doi.org/10.1080/19315775.2007.11721402} {\bibfield  {journal} {\bibinfo  {journal} {NCSLI Measure}\ }\textbf {\bibinfo {volume} {2}},\ \bibinfo {pages} {74–89} (\bibinfo {year} {2007})}\BibitemShut {NoStop}%
\bibitem [{\citenamefont {Tierney}\ \emph {et~al.}(2019)\citenamefont {Tierney}, \citenamefont {Holmes}, \citenamefont {Mellor}, \citenamefont {López}, \citenamefont {Roberts}, \citenamefont {Hill}, \citenamefont {Boto}, \citenamefont {Leggett}, \citenamefont {Shah}, \citenamefont {Brookes}, \citenamefont {Bowtell},\ and\ \citenamefont {Barnes}}]{Tierney2019OPMreview}%
  \BibitemOpen
  \bibfield  {author} {\bibinfo {author} {\bibfnamefont {T.~M.}\ \bibnamefont {Tierney}}, \bibinfo {author} {\bibfnamefont {N.}~\bibnamefont {Holmes}}, \bibinfo {author} {\bibfnamefont {S.}~\bibnamefont {Mellor}}, \bibinfo {author} {\bibfnamefont {J.~D.}\ \bibnamefont {López}}, \bibinfo {author} {\bibfnamefont {G.}~\bibnamefont {Roberts}}, \bibinfo {author} {\bibfnamefont {R.~M.}\ \bibnamefont {Hill}}, \bibinfo {author} {\bibfnamefont {E.}~\bibnamefont {Boto}}, \bibinfo {author} {\bibfnamefont {J.}~\bibnamefont {Leggett}}, \bibinfo {author} {\bibfnamefont {V.}~\bibnamefont {Shah}}, \bibinfo {author} {\bibfnamefont {M.~J.}\ \bibnamefont {Brookes}}, \bibinfo {author} {\bibfnamefont {R.}~\bibnamefont {Bowtell}},\ and\ \bibinfo {author} {\bibfnamefont {G.~R.}\ \bibnamefont {Barnes}},\ }\bibfield  {title} {\enquote {\bibinfo {title} {Optically pumped magnetometers: From quantum origins to multi-channel magnetoencephalography},}\ }\href {https://doi.org/10.1016/j.neuroimage.2019.05.063} {\bibfield  {journal}
  {\bibinfo  {journal} {NeuroImage}\ }\textbf {\bibinfo {volume} {199}},\ \bibinfo {pages} {598–608} (\bibinfo {year} {2019})}\BibitemShut {NoStop}%
\bibitem [{\citenamefont {Degen}, \citenamefont {Reinhard},\ and\ \citenamefont {Cappellaro}(2017)}]{RevModPhys.89.035002}%
  \BibitemOpen
  \bibfield  {author} {\bibinfo {author} {\bibfnamefont {C.~L.}\ \bibnamefont {Degen}}, \bibinfo {author} {\bibfnamefont {F.}~\bibnamefont {Reinhard}},\ and\ \bibinfo {author} {\bibfnamefont {P.}~\bibnamefont {Cappellaro}},\ }\bibfield  {title} {\enquote {\bibinfo {title} {Quantum sensing},}\ }\href {https://doi.org/10.1103/RevModPhys.89.035002} {\bibfield  {journal} {\bibinfo  {journal} {Rev. Mod. Phys.}\ }\textbf {\bibinfo {volume} {89}},\ \bibinfo {pages} {035002} (\bibinfo {year} {2017})}\BibitemShut {NoStop}%
\bibitem [{\citenamefont {Knappe}\ \emph {et~al.}(2004)\citenamefont {Knappe}, \citenamefont {Shah}, \citenamefont {Schwindt}, \citenamefont {Hollberg}, \citenamefont {Kitching}, \citenamefont {Liew},\ and\ \citenamefont {Moreland}}]{KnappeShahKitching2}%
  \BibitemOpen
  \bibfield  {author} {\bibinfo {author} {\bibfnamefont {S.}~\bibnamefont {Knappe}}, \bibinfo {author} {\bibfnamefont {V.}~\bibnamefont {Shah}}, \bibinfo {author} {\bibfnamefont {P.~D.~D.}\ \bibnamefont {Schwindt}}, \bibinfo {author} {\bibfnamefont {L.}~\bibnamefont {Hollberg}}, \bibinfo {author} {\bibfnamefont {J.}~\bibnamefont {Kitching}}, \bibinfo {author} {\bibfnamefont {L.-A.}\ \bibnamefont {Liew}},\ and\ \bibinfo {author} {\bibfnamefont {J.}~\bibnamefont {Moreland}},\ }\bibfield  {title} {\enquote {\bibinfo {title} {{A microfabricated atomic clock}},}\ }\href {https://doi.org/10.1063/1.1787942} {\bibfield  {journal} {\bibinfo  {journal} {Applied Physics Letters}\ }\textbf {\bibinfo {volume} {85}},\ \bibinfo {pages} {1460--1462} (\bibinfo {year} {2004})}\BibitemShut {NoStop}%
\bibitem [{\citenamefont {Liew}\ \emph {et~al.}(2004)\citenamefont {Liew}, \citenamefont {Knappe}, \citenamefont {Moreland}, \citenamefont {Robinson}, \citenamefont {Hollberg},\ and\ \citenamefont {Kitching}}]{liew2004microfabricated}%
  \BibitemOpen
  \bibfield  {author} {\bibinfo {author} {\bibfnamefont {L.-A.}\ \bibnamefont {Liew}}, \bibinfo {author} {\bibfnamefont {S.}~\bibnamefont {Knappe}}, \bibinfo {author} {\bibfnamefont {J.}~\bibnamefont {Moreland}}, \bibinfo {author} {\bibfnamefont {H.}~\bibnamefont {Robinson}}, \bibinfo {author} {\bibfnamefont {L.}~\bibnamefont {Hollberg}},\ and\ \bibinfo {author} {\bibfnamefont {J.}~\bibnamefont {Kitching}},\ }\bibfield  {title} {\enquote {\bibinfo {title} {Microfabricated alkali atom vapor cells},}\ }\href@noop {} {\bibfield  {journal} {\bibinfo  {journal} {Applied Physics Letters}\ }\textbf {\bibinfo {volume} {84}},\ \bibinfo {pages} {2694--2696} (\bibinfo {year} {2004})}\BibitemShut {NoStop}%
\bibitem [{\citenamefont {Schwindt}\ \emph {et~al.}(2004)\citenamefont {Schwindt}, \citenamefont {Knappe}, \citenamefont {Shah}, \citenamefont {Hollberg}, \citenamefont {Kitching}, \citenamefont {Liew},\ and\ \citenamefont {Moreland}}]{KnappeShahKitching}%
  \BibitemOpen
  \bibfield  {author} {\bibinfo {author} {\bibfnamefont {P.~D.~D.}\ \bibnamefont {Schwindt}}, \bibinfo {author} {\bibfnamefont {S.}~\bibnamefont {Knappe}}, \bibinfo {author} {\bibfnamefont {V.}~\bibnamefont {Shah}}, \bibinfo {author} {\bibfnamefont {L.}~\bibnamefont {Hollberg}}, \bibinfo {author} {\bibfnamefont {J.}~\bibnamefont {Kitching}}, \bibinfo {author} {\bibfnamefont {L.-A.}\ \bibnamefont {Liew}},\ and\ \bibinfo {author} {\bibfnamefont {J.}~\bibnamefont {Moreland}},\ }\bibfield  {title} {\enquote {\bibinfo {title} {{Chip-scale atomic magnetometer}},}\ }\href {https://doi.org/10.1063/1.1839274} {\bibfield  {journal} {\bibinfo  {journal} {Applied Physics Letters}\ }\textbf {\bibinfo {volume} {85}},\ \bibinfo {pages} {6409--6411} (\bibinfo {year} {2004})}\BibitemShut {NoStop}%
\bibitem [{\citenamefont {Kitching}, \citenamefont {Knappe},\ and\ \citenamefont {Donley}(2011)}]{kitching2011atomic}%
  \BibitemOpen
  \bibfield  {author} {\bibinfo {author} {\bibfnamefont {J.}~\bibnamefont {Kitching}}, \bibinfo {author} {\bibfnamefont {S.}~\bibnamefont {Knappe}},\ and\ \bibinfo {author} {\bibfnamefont {E.~A.}\ \bibnamefont {Donley}},\ }\bibfield  {title} {\enquote {\bibinfo {title} {Atomic sensors--a review},}\ }\href@noop {} {\bibfield  {journal} {\bibinfo  {journal} {IEEE Sensors Journal}\ }\textbf {\bibinfo {volume} {11}},\ \bibinfo {pages} {1749--1758} (\bibinfo {year} {2011})}\BibitemShut {NoStop}%
\bibitem [{\citenamefont {Kitching}(2018)}]{kitchingReview}%
  \BibitemOpen
  \bibfield  {author} {\bibinfo {author} {\bibfnamefont {J.}~\bibnamefont {Kitching}},\ }\bibfield  {title} {\enquote {\bibinfo {title} {{Chip-scale atomic devices}},}\ }\href {https://doi.org/10.1063/1.5026238} {\bibfield  {journal} {\bibinfo  {journal} {Applied Physics Reviews}\ }\textbf {\bibinfo {volume} {5}} (\bibinfo {year} {2018}),\ 10.1063/1.5026238}\BibitemShut {NoStop}%
\bibitem [{\citenamefont {Haesler}\ \emph {et~al.}(2017)\citenamefont {Haesler}, \citenamefont {Balet}, \citenamefont {Karlen}, \citenamefont {Overstolz}, \citenamefont {Gallinet}, \citenamefont {Lecomte}, \citenamefont {Droz}, \citenamefont {Kautio}, \citenamefont {Karioja}, \citenamefont {Lahti}, \citenamefont {Määttänen}, \citenamefont {Lahtinen},\ and\ \citenamefont {Hevonkorpi}}]{Haesler2017}%
  \BibitemOpen
  \bibfield  {author} {\bibinfo {author} {\bibfnamefont {J.}~\bibnamefont {Haesler}}, \bibinfo {author} {\bibfnamefont {L.}~\bibnamefont {Balet}}, \bibinfo {author} {\bibfnamefont {S.}~\bibnamefont {Karlen}}, \bibinfo {author} {\bibfnamefont {T.}~\bibnamefont {Overstolz}}, \bibinfo {author} {\bibfnamefont {B.}~\bibnamefont {Gallinet}}, \bibinfo {author} {\bibfnamefont {S.}~\bibnamefont {Lecomte}}, \bibinfo {author} {\bibfnamefont {F.}~\bibnamefont {Droz}}, \bibinfo {author} {\bibfnamefont {K.}~\bibnamefont {Kautio}}, \bibinfo {author} {\bibfnamefont {P.}~\bibnamefont {Karioja}}, \bibinfo {author} {\bibfnamefont {M.}~\bibnamefont {Lahti}}, \bibinfo {author} {\bibfnamefont {A.}~\bibnamefont {Määttänen}}, \bibinfo {author} {\bibfnamefont {O.}~\bibnamefont {Lahtinen}},\ and\ \bibinfo {author} {\bibfnamefont {V.}~\bibnamefont {Hevonkorpi}},\ }\bibfield  {title} {\enquote {\bibinfo {title} {Low-power and low-profile miniature atomic clock ceramic based flat form factor miniature atomic clock physics package
  ({C-MAC})},}\ }in\ \href@noop {} {\emph {\bibinfo {booktitle} {European Frequency and Time Forum \& International Frequency Control Symposium (EFTF-IFCS)}}}\ (\bibinfo {year} {2017})\BibitemShut {NoStop}%
\bibitem [{\citenamefont {Boto}\ \emph {et~al.}(2018)\citenamefont {Boto}, \citenamefont {Holmes}, \citenamefont {Leggett}, \citenamefont {Roberts}, \citenamefont {Shah}, \citenamefont {Meyer}, \citenamefont {Mu{\~{n}}oz}, \citenamefont {Mullinger}, \citenamefont {Tierney}, \citenamefont {Bestmann}, \citenamefont {Barnes}, \citenamefont {Bowtell},\ and\ \citenamefont {Brookes}}]{Boto2018}%
  \BibitemOpen
  \bibfield  {author} {\bibinfo {author} {\bibfnamefont {E.}~\bibnamefont {Boto}}, \bibinfo {author} {\bibfnamefont {N.}~\bibnamefont {Holmes}}, \bibinfo {author} {\bibfnamefont {J.}~\bibnamefont {Leggett}}, \bibinfo {author} {\bibfnamefont {G.}~\bibnamefont {Roberts}}, \bibinfo {author} {\bibfnamefont {V.}~\bibnamefont {Shah}}, \bibinfo {author} {\bibfnamefont {S.~S.}\ \bibnamefont {Meyer}}, \bibinfo {author} {\bibfnamefont {L.~D.}\ \bibnamefont {Mu{\~{n}}oz}}, \bibinfo {author} {\bibfnamefont {K.~J.}\ \bibnamefont {Mullinger}}, \bibinfo {author} {\bibfnamefont {T.~M.}\ \bibnamefont {Tierney}}, \bibinfo {author} {\bibfnamefont {S.}~\bibnamefont {Bestmann}}, \bibinfo {author} {\bibfnamefont {G.~R.}\ \bibnamefont {Barnes}}, \bibinfo {author} {\bibfnamefont {R.}~\bibnamefont {Bowtell}},\ and\ \bibinfo {author} {\bibfnamefont {M.~J.}\ \bibnamefont {Brookes}},\ }\bibfield  {title} {\enquote {\bibinfo {title} {Moving magnetoencephalography towards real-world applications with a wearable system},}\ }\href
  {https://doi.org/10.1038/nature26147} {\bibfield  {journal} {\bibinfo  {journal} {Nature}\ }\textbf {\bibinfo {volume} {555}},\ \bibinfo {pages} {657--661} (\bibinfo {year} {2018})}\BibitemShut {NoStop}%
\bibitem [{\citenamefont {MacFarlane}, \citenamefont {Dowling},\ and\ \citenamefont {Milburn}(2003)}]{MacFarlane2003RSTA361}%
  \BibitemOpen
  \bibfield  {author} {\bibinfo {author} {\bibfnamefont {A.~G.~J.}\ \bibnamefont {MacFarlane}}, \bibinfo {author} {\bibfnamefont {J.~P.}\ \bibnamefont {Dowling}},\ and\ \bibinfo {author} {\bibfnamefont {G.~J.}\ \bibnamefont {Milburn}},\ }\bibfield  {title} {\enquote {\bibinfo {title} {Quantum technology: the second quantum revolution},}\ }\href {https://doi.org/10.1098/rsta.2003.1227} {\bibfield  {journal} {\bibinfo  {journal} {Philosophical Transactions of the Royal Society of London. Series A: Mathematical, Physical and Engineering Sciences}\ }\textbf {\bibinfo {volume} {361}},\ \bibinfo {pages} {1655--1674} (\bibinfo {year} {2003})}\BibitemShut {NoStop}%
\bibitem [{\citenamefont {Lipka}\ \emph {et~al.}(2024)\citenamefont {Lipka}, \citenamefont {Sierant}, \citenamefont {Troullinou},\ and\ \citenamefont {Mitchell}}]{Lipka2024multiparameter}%
  \BibitemOpen
  \bibfield  {author} {\bibinfo {author} {\bibfnamefont {M.}~\bibnamefont {Lipka}}, \bibinfo {author} {\bibfnamefont {A.}~\bibnamefont {Sierant}}, \bibinfo {author} {\bibfnamefont {C.}~\bibnamefont {Troullinou}},\ and\ \bibinfo {author} {\bibfnamefont {M.~W.}\ \bibnamefont {Mitchell}},\ }\bibfield  {title} {\enquote {\bibinfo {title} {Multiparameter quantum sensing and magnetic communication with a hybrid dc and rf optically pumped magnetometer},}\ }\href {https://doi.org/10.1103/physrevapplied.21.034054} {\bibfield  {journal} {\bibinfo  {journal} {Physical Review Applied}\ }\textbf {\bibinfo {volume} {21}} (\bibinfo {year} {2024}),\ 10.1103/physrevapplied.21.034054}\BibitemShut {NoStop}%
\bibitem [{\citenamefont {Ledbetter}\ \emph {et~al.}(2008)\citenamefont {Ledbetter}, \citenamefont {Savukov}, \citenamefont {Budker}, \citenamefont {Shah}, \citenamefont {Knappe}, \citenamefont {Kitching}, \citenamefont {Michalak}, \citenamefont {Xu},\ and\ \citenamefont {Pines}}]{Ledbetter2008PNAS}%
  \BibitemOpen
  \bibfield  {author} {\bibinfo {author} {\bibfnamefont {M.~P.}\ \bibnamefont {Ledbetter}}, \bibinfo {author} {\bibfnamefont {I.~M.}\ \bibnamefont {Savukov}}, \bibinfo {author} {\bibfnamefont {D.}~\bibnamefont {Budker}}, \bibinfo {author} {\bibfnamefont {V.}~\bibnamefont {Shah}}, \bibinfo {author} {\bibfnamefont {S.}~\bibnamefont {Knappe}}, \bibinfo {author} {\bibfnamefont {J.}~\bibnamefont {Kitching}}, \bibinfo {author} {\bibfnamefont {D.~J.}\ \bibnamefont {Michalak}}, \bibinfo {author} {\bibfnamefont {S.}~\bibnamefont {Xu}},\ and\ \bibinfo {author} {\bibfnamefont {A.}~\bibnamefont {Pines}},\ }\bibfield  {title} {\enquote {\bibinfo {title} {Zero-field remote detection of {NMR} with a microfabricated atomic magnetometer},}\ }\href {https://doi.org/10.1073/pnas.0711505105} {\bibfield  {journal} {\bibinfo  {journal} {Proceedings of the National Academy of Sciences}\ }\textbf {\bibinfo {volume} {105}},\ \bibinfo {pages} {2286–2290} (\bibinfo {year} {2008})}\BibitemShut {NoStop}%
\bibitem [{\citenamefont {Yu}, \citenamefont {Garcia},\ and\ \citenamefont {Xu}(2009)}]{Yu2009CMRA}%
  \BibitemOpen
  \bibfield  {author} {\bibinfo {author} {\bibfnamefont {D.}~\bibnamefont {Yu}}, \bibinfo {author} {\bibfnamefont {N.}~\bibnamefont {Garcia}},\ and\ \bibinfo {author} {\bibfnamefont {S.}~\bibnamefont {Xu}},\ }\bibfield  {title} {\enquote {\bibinfo {title} {Toward portable nuclear magnetic resonance devices using atomic magnetometers},}\ }\href {https://doi.org/10.1002/cmr.a.20134} {\bibfield  {journal} {\bibinfo  {journal} {Concepts in Magnetic Resonance Part A}\ }\textbf {\bibinfo {volume} {34A}},\ \bibinfo {pages} {124–132} (\bibinfo {year} {2009})}\BibitemShut {NoStop}%
\bibitem [{\citenamefont {Blanchard}\ \emph {et~al.}(2020)\citenamefont {Blanchard}, \citenamefont {Wu}, \citenamefont {Eills}, \citenamefont {Hu},\ and\ \citenamefont {Budker}}]{Blanchard2020JMR}%
  \BibitemOpen
  \bibfield  {author} {\bibinfo {author} {\bibfnamefont {J.~W.}\ \bibnamefont {Blanchard}}, \bibinfo {author} {\bibfnamefont {T.}~\bibnamefont {Wu}}, \bibinfo {author} {\bibfnamefont {J.}~\bibnamefont {Eills}}, \bibinfo {author} {\bibfnamefont {Y.}~\bibnamefont {Hu}},\ and\ \bibinfo {author} {\bibfnamefont {D.}~\bibnamefont {Budker}},\ }\bibfield  {title} {\enquote {\bibinfo {title} {Zero- to ultralow-field nuclear magnetic resonance {J}-spectroscopy with commercial atomic magnetometers},}\ }\href {https://doi.org/10.1016/j.jmr.2020.106723} {\bibfield  {journal} {\bibinfo  {journal} {Journal of Magnetic Resonance}\ }\textbf {\bibinfo {volume} {314}},\ \bibinfo {pages} {106723} (\bibinfo {year} {2020})}\BibitemShut {NoStop}%
\bibitem [{\citenamefont {Mouloudakis}\ \emph {et~al.}(2023)\citenamefont {Mouloudakis}, \citenamefont {Bodenstedt}, \citenamefont {Azagra}, \citenamefont {Mitchell}, \citenamefont {Marco-Rius},\ and\ \citenamefont {Tayler}}]{Mouloudakis2023JPCL}%
  \BibitemOpen
  \bibfield  {author} {\bibinfo {author} {\bibfnamefont {K.}~\bibnamefont {Mouloudakis}}, \bibinfo {author} {\bibfnamefont {S.}~\bibnamefont {Bodenstedt}}, \bibinfo {author} {\bibfnamefont {M.}~\bibnamefont {Azagra}}, \bibinfo {author} {\bibfnamefont {M.~W.}\ \bibnamefont {Mitchell}}, \bibinfo {author} {\bibfnamefont {I.}~\bibnamefont {Marco-Rius}},\ and\ \bibinfo {author} {\bibfnamefont {M.~C.~D.}\ \bibnamefont {Tayler}},\ }\bibfield  {title} {\enquote {\bibinfo {title} {Real-time polarimetry of hyperpolarized $^{13}${C} nuclear spins using an atomic magnetometer},}\ }\href {https://doi.org/10.1021/acs.jpclett.2c03864} {\bibfield  {journal} {\bibinfo  {journal} {The Journal of Physical Chemistry Letters}\ }\textbf {\bibinfo {volume} {14}},\ \bibinfo {pages} {1192--1197} (\bibinfo {year} {2023})}\BibitemShut {NoStop}%
\bibitem [{\citenamefont {Maurice}\ \emph {et~al.}(2022)\citenamefont {Maurice}, \citenamefont {Carl{\'{e}}}, \citenamefont {Keshavarzi}, \citenamefont {Chutani}, \citenamefont {Queste}, \citenamefont {Gauthier-Manuel}, \citenamefont {Cote}, \citenamefont {Vicarini}, \citenamefont {Hafiz}, \citenamefont {Boudot},\ and\ \citenamefont {Passilly}}]{Maurice2022}%
  \BibitemOpen
  \bibfield  {author} {\bibinfo {author} {\bibfnamefont {V.}~\bibnamefont {Maurice}}, \bibinfo {author} {\bibfnamefont {C.}~\bibnamefont {Carl{\'{e}}}}, \bibinfo {author} {\bibfnamefont {S.}~\bibnamefont {Keshavarzi}}, \bibinfo {author} {\bibfnamefont {R.}~\bibnamefont {Chutani}}, \bibinfo {author} {\bibfnamefont {S.}~\bibnamefont {Queste}}, \bibinfo {author} {\bibfnamefont {L.}~\bibnamefont {Gauthier-Manuel}}, \bibinfo {author} {\bibfnamefont {J.-M.}\ \bibnamefont {Cote}}, \bibinfo {author} {\bibfnamefont {R.}~\bibnamefont {Vicarini}}, \bibinfo {author} {\bibfnamefont {M.~A.}\ \bibnamefont {Hafiz}}, \bibinfo {author} {\bibfnamefont {R.}~\bibnamefont {Boudot}},\ and\ \bibinfo {author} {\bibfnamefont {N.}~\bibnamefont {Passilly}},\ }\bibfield  {title} {\enquote {\bibinfo {title} {Wafer-level vapor cells filled with laser-actuated hermetic seals for integrated atomic devices},}\ }\href {https://doi.org/10.1038/s41378-022-00468-x} {\bibfield  {journal} {\bibinfo  {journal} {Microsystems $\&$ Nanoengineering}\
  }\textbf {\bibinfo {volume} {8}} (\bibinfo {year} {2022}),\ 10.1038/s41378-022-00468-x}\BibitemShut {NoStop}%
\bibitem [{\citenamefont {Sheng}\ \emph {et~al.}(2013)\citenamefont {Sheng}, \citenamefont {Li}, \citenamefont {Dural},\ and\ \citenamefont {Romalis}}]{PhysRevLett.110.160802}%
  \BibitemOpen
  \bibfield  {author} {\bibinfo {author} {\bibfnamefont {D.}~\bibnamefont {Sheng}}, \bibinfo {author} {\bibfnamefont {S.}~\bibnamefont {Li}}, \bibinfo {author} {\bibfnamefont {N.}~\bibnamefont {Dural}},\ and\ \bibinfo {author} {\bibfnamefont {M.~V.}\ \bibnamefont {Romalis}},\ }\bibfield  {title} {\enquote {\bibinfo {title} {Subfemtotesla scalar atomic magnetometry using multipass cells},}\ }\href {https://doi.org/10.1103/PhysRevLett.110.160802} {\bibfield  {journal} {\bibinfo  {journal} {Phys. Rev. Lett.}\ }\textbf {\bibinfo {volume} {110}},\ \bibinfo {pages} {160802} (\bibinfo {year} {2013})}\BibitemShut {NoStop}%
\bibitem [{\citenamefont {Yu}\ \emph {et~al.}(2023)\citenamefont {Yu}, \citenamefont {Chen}, \citenamefont {Wang}, \citenamefont {Han}, \citenamefont {Luo}, \citenamefont {Zhao}, \citenamefont {Wang}, \citenamefont {Ma}, \citenamefont {Lu}, \citenamefont {Yang}, \citenamefont {Lin}, \citenamefont {Wang},\ and\ \citenamefont {Jiang}}]{Yu2023}%
  \BibitemOpen
  \bibfield  {author} {\bibinfo {author} {\bibfnamefont {M.}~\bibnamefont {Yu}}, \bibinfo {author} {\bibfnamefont {Y.}~\bibnamefont {Chen}}, \bibinfo {author} {\bibfnamefont {Y.}~\bibnamefont {Wang}}, \bibinfo {author} {\bibfnamefont {X.}~\bibnamefont {Han}}, \bibinfo {author} {\bibfnamefont {G.}~\bibnamefont {Luo}}, \bibinfo {author} {\bibfnamefont {L.}~\bibnamefont {Zhao}}, \bibinfo {author} {\bibfnamefont {Y.}~\bibnamefont {Wang}}, \bibinfo {author} {\bibfnamefont {Y.}~\bibnamefont {Ma}}, \bibinfo {author} {\bibfnamefont {S.}~\bibnamefont {Lu}}, \bibinfo {author} {\bibfnamefont {P.}~\bibnamefont {Yang}}, \bibinfo {author} {\bibfnamefont {Q.}~\bibnamefont {Lin}}, \bibinfo {author} {\bibfnamefont {K.}~\bibnamefont {Wang}},\ and\ \bibinfo {author} {\bibfnamefont {Z.}~\bibnamefont {Jiang}},\ }\bibfield  {title} {\enquote {\bibinfo {title} {Microfabricated atomic vapor cells with multi-optical channels based on an innovative inner-sidewall molding process},}\ }\href {https://doi.org/10.1016/j.eng.2023.08.016}
  {\bibfield  {journal} {\bibinfo  {journal} {Engineering}\ ,\ \bibinfo {pages} {In press}} (\bibinfo {year} {2023})}\BibitemShut {NoStop}%
\bibitem [{\citenamefont {Wu}(2021)}]{RevModPhys.93.035006}%
  \BibitemOpen
  \bibfield  {author} {\bibinfo {author} {\bibfnamefont {Z.}~\bibnamefont {Wu}},\ }\bibfield  {title} {\enquote {\bibinfo {title} {Wall interactions of spin-polarized atoms},}\ }\href {https://doi.org/10.1103/RevModPhys.93.035006} {\bibfield  {journal} {\bibinfo  {journal} {Rev. Mod. Phys.}\ }\textbf {\bibinfo {volume} {93}},\ \bibinfo {pages} {035006} (\bibinfo {year} {2021})}\BibitemShut {NoStop}%
\bibitem [{\citenamefont {Dyer}\ \emph {et~al.}(2022)\citenamefont {Dyer}, \citenamefont {Griffin}, \citenamefont {Arnold}, \citenamefont {Mirando}, \citenamefont {Burt}, \citenamefont {Riis},\ and\ \citenamefont {McGilligan}}]{Dyer2022}%
  \BibitemOpen
  \bibfield  {author} {\bibinfo {author} {\bibfnamefont {S.}~\bibnamefont {Dyer}}, \bibinfo {author} {\bibfnamefont {P.~F.}\ \bibnamefont {Griffin}}, \bibinfo {author} {\bibfnamefont {A.~S.}\ \bibnamefont {Arnold}}, \bibinfo {author} {\bibfnamefont {F.}~\bibnamefont {Mirando}}, \bibinfo {author} {\bibfnamefont {D.~P.}\ \bibnamefont {Burt}}, \bibinfo {author} {\bibfnamefont {E.}~\bibnamefont {Riis}},\ and\ \bibinfo {author} {\bibfnamefont {J.~P.}\ \bibnamefont {McGilligan}},\ }\bibfield  {title} {\enquote {\bibinfo {title} {Micro-machined deep silicon atomic vapor cells},}\ }\href {https://doi.org/10.1063/5.0114762} {\bibfield  {journal} {\bibinfo  {journal} {Journal of Applied Physics}\ }\textbf {\bibinfo {volume} {132}},\ \bibinfo {pages} {134401} (\bibinfo {year} {2022})}\BibitemShut {NoStop}%
\bibitem [{\citenamefont {Gallinet}\ \emph {et~al.}(2019)\citenamefont {Gallinet}, \citenamefont {Haesler}, \citenamefont {Lecomte},\ and\ \citenamefont {Basset}}]{gallinet2019atomic}%
  \BibitemOpen
  \bibfield  {author} {\bibinfo {author} {\bibfnamefont {B.}~\bibnamefont {Gallinet}}, \bibinfo {author} {\bibfnamefont {J.}~\bibnamefont {Haesler}}, \bibinfo {author} {\bibfnamefont {S.}~\bibnamefont {Lecomte}},\ and\ \bibinfo {author} {\bibfnamefont {G.}~\bibnamefont {Basset}},\ }\href@noop {} {\enquote {\bibinfo {title} {Atomic clock},}\ } (\bibinfo {year} {2019}),\ \bibinfo {note} {{US Patent 10,423,124}}\BibitemShut {NoStop}%
\bibitem [{\citenamefont {Knappe}(2007)}]{Knappe220921}%
  \BibitemOpen
  \bibfield  {author} {\bibinfo {author} {\bibfnamefont {S.}~\bibnamefont {Knappe}},\ }\href {https://tsapps.nist.gov/publication/get_pdf.cfm?pub_id=50424} {\emph {\bibinfo {title} {Emerging Topics: MEMS Atomic Clocks}}},\ \bibinfo {number} {3}\ (\bibinfo  {publisher} {Elsevier (Netherlands), NL},\ \bibinfo {year} {2007})\BibitemShut {NoStop}%
\bibitem [{\citenamefont {Lutwak}\ \emph {et~al.}(2007)\citenamefont {Lutwak}, \citenamefont {Rashed}, \citenamefont {Varghese}, \citenamefont {Tepolt}, \citenamefont {Leblanc}, \citenamefont {Mescher}, \citenamefont {Serkland},\ and\ \citenamefont {Peake}}]{Lutwak4319292}%
  \BibitemOpen
  \bibfield  {author} {\bibinfo {author} {\bibfnamefont {R.}~\bibnamefont {Lutwak}}, \bibinfo {author} {\bibfnamefont {A.}~\bibnamefont {Rashed}}, \bibinfo {author} {\bibfnamefont {M.}~\bibnamefont {Varghese}}, \bibinfo {author} {\bibfnamefont {G.}~\bibnamefont {Tepolt}}, \bibinfo {author} {\bibfnamefont {J.}~\bibnamefont {Leblanc}}, \bibinfo {author} {\bibfnamefont {M.}~\bibnamefont {Mescher}}, \bibinfo {author} {\bibfnamefont {D.}~\bibnamefont {Serkland}},\ and\ \bibinfo {author} {\bibfnamefont {G.}~\bibnamefont {Peake}},\ }\bibfield  {title} {\enquote {\bibinfo {title} {The miniature atomic clock - pre-production results},}\ }in\ \href {https://doi.org/10.1109/FREQ.2007.4319292} {\emph {\bibinfo {booktitle} {2007 IEEE International Frequency Control Symposium Joint with the 21st European Frequency and Time Forum}}}\ (\bibinfo {year} {2007})\ pp.\ \bibinfo {pages} {1327--1333}\BibitemShut {NoStop}%
\bibitem [{\citenamefont {Affolderbach}\ \emph {et~al.}(2015)\citenamefont {Affolderbach}, \citenamefont {Du}, \citenamefont {Bandi}, \citenamefont {Horsley}, \citenamefont {Treutlein},\ and\ \citenamefont {Mileti}}]{Affolderbach7140794}%
  \BibitemOpen
  \bibfield  {author} {\bibinfo {author} {\bibfnamefont {C.}~\bibnamefont {Affolderbach}}, \bibinfo {author} {\bibfnamefont {G.-X.}\ \bibnamefont {Du}}, \bibinfo {author} {\bibfnamefont {T.}~\bibnamefont {Bandi}}, \bibinfo {author} {\bibfnamefont {A.}~\bibnamefont {Horsley}}, \bibinfo {author} {\bibfnamefont {P.}~\bibnamefont {Treutlein}},\ and\ \bibinfo {author} {\bibfnamefont {G.}~\bibnamefont {Mileti}},\ }\bibfield  {title} {\enquote {\bibinfo {title} {Imaging microwave and {DC} magnetic fields in a vapor-cell {Rb} atomic clock},}\ }\href {https://doi.org/10.1109/TIM.2015.2444261} {\bibfield  {journal} {\bibinfo  {journal} {IEEE Transactions on Instrumentation and Measurement}\ }\textbf {\bibinfo {volume} {64}},\ \bibinfo {pages} {3629--3637} (\bibinfo {year} {2015})}\BibitemShut {NoStop}%
\bibitem [{\citenamefont {P\'etremand}\ \emph {et~al.}(2012)\citenamefont {P\'etremand}, \citenamefont {Affolderbach}, \citenamefont {Straessle}, \citenamefont {Pellaton}, \citenamefont {Briand}, \citenamefont {Mileti},\ and\ \citenamefont {de~Rooij}}]{Petremand2012}%
  \BibitemOpen
  \bibfield  {author} {\bibinfo {author} {\bibfnamefont {Y.}~\bibnamefont {P\'etremand}}, \bibinfo {author} {\bibfnamefont {C.}~\bibnamefont {Affolderbach}}, \bibinfo {author} {\bibfnamefont {R.}~\bibnamefont {Straessle}}, \bibinfo {author} {\bibfnamefont {M.}~\bibnamefont {Pellaton}}, \bibinfo {author} {\bibfnamefont {D.}~\bibnamefont {Briand}}, \bibinfo {author} {\bibfnamefont {G.}~\bibnamefont {Mileti}},\ and\ \bibinfo {author} {\bibfnamefont {N.~F.}\ \bibnamefont {de~Rooij}},\ }\bibfield  {title} {\enquote {\bibinfo {title} {Microfabricated rubidium vapour cell with a thick glass core for small-scale atomic clock applications},}\ }\href {https://doi.org/10.1088/0960-1317/22/2/025013} {\bibfield  {journal} {\bibinfo  {journal} {Journal of Micromechanics and Microengineering}\ }\textbf {\bibinfo {volume} {22}},\ \bibinfo {pages} {025013} (\bibinfo {year} {2012})}\BibitemShut {NoStop}%
\bibitem [{\citenamefont {Dyer}\ \emph {et~al.}(2023)\citenamefont {Dyer}, \citenamefont {McWilliam}, \citenamefont {Hunter}, \citenamefont {Ingleby}, \citenamefont {Burt}, \citenamefont {Sharp}, \citenamefont {Mirando}, \citenamefont {Griffin}, \citenamefont {Riis},\ and\ \citenamefont {McGilligan}}]{dyer2023real}%
  \BibitemOpen
  \bibfield  {author} {\bibinfo {author} {\bibfnamefont {S.}~\bibnamefont {Dyer}}, \bibinfo {author} {\bibfnamefont {A.}~\bibnamefont {McWilliam}}, \bibinfo {author} {\bibfnamefont {D.}~\bibnamefont {Hunter}}, \bibinfo {author} {\bibfnamefont {S.}~\bibnamefont {Ingleby}}, \bibinfo {author} {\bibfnamefont {D.}~\bibnamefont {Burt}}, \bibinfo {author} {\bibfnamefont {O.}~\bibnamefont {Sharp}}, \bibinfo {author} {\bibfnamefont {F.}~\bibnamefont {Mirando}}, \bibinfo {author} {\bibfnamefont {P.}~\bibnamefont {Griffin}}, \bibinfo {author} {\bibfnamefont {E.}~\bibnamefont {Riis}},\ and\ \bibinfo {author} {\bibfnamefont {J.}~\bibnamefont {McGilligan}},\ }\href@noop {} {\enquote {\bibinfo {title} {Real-time buffer gas pressure tuning in a micro-machined vapor cell},}\ } (\bibinfo {year} {2023}),\ \Eprint {https://arxiv.org/abs/2304.06153} {arXiv:2304.06153 [atom-ph]} \BibitemShut {NoStop}%
\bibitem [{\citenamefont {Martinez}\ \emph {et~al.}(2023)\citenamefont {Martinez}, \citenamefont {Li}, \citenamefont {Staron}, \citenamefont {Kitching}, \citenamefont {Raman},\ and\ \citenamefont {McGehee}}]{Martinez2023}%
  \BibitemOpen
  \bibfield  {author} {\bibinfo {author} {\bibfnamefont {G.~D.}\ \bibnamefont {Martinez}}, \bibinfo {author} {\bibfnamefont {C.}~\bibnamefont {Li}}, \bibinfo {author} {\bibfnamefont {A.}~\bibnamefont {Staron}}, \bibinfo {author} {\bibfnamefont {J.}~\bibnamefont {Kitching}}, \bibinfo {author} {\bibfnamefont {C.}~\bibnamefont {Raman}},\ and\ \bibinfo {author} {\bibfnamefont {W.~R.}\ \bibnamefont {McGehee}},\ }\bibfield  {title} {\enquote {\bibinfo {title} {A chip-scale atomic beam clock},}\ }\href {https://doi.org/10.1038/s41467-023-39166-1} {\bibfield  {journal} {\bibinfo  {journal} {Nature Communications}\ }\textbf {\bibinfo {volume} {14}} (\bibinfo {year} {2023}),\ 10.1038/s41467-023-39166-1}\BibitemShut {NoStop}%
\bibitem [{\citenamefont {Batori}\ \emph {et~al.}(2022)\citenamefont {Batori}, \citenamefont {Affolderbach}, \citenamefont {Pellaton}, \citenamefont {Gruet}, \citenamefont {Violetti}, \citenamefont {Su}, \citenamefont {Skrivervik},\ and\ \citenamefont {Mileti}}]{PhysRevApplied.18.054039}%
  \BibitemOpen
  \bibfield  {author} {\bibinfo {author} {\bibfnamefont {E.}~\bibnamefont {Batori}}, \bibinfo {author} {\bibfnamefont {C.}~\bibnamefont {Affolderbach}}, \bibinfo {author} {\bibfnamefont {M.}~\bibnamefont {Pellaton}}, \bibinfo {author} {\bibfnamefont {F.}~\bibnamefont {Gruet}}, \bibinfo {author} {\bibfnamefont {M.}~\bibnamefont {Violetti}}, \bibinfo {author} {\bibfnamefont {Y.}~\bibnamefont {Su}}, \bibinfo {author} {\bibfnamefont {A.~K.}\ \bibnamefont {Skrivervik}},\ and\ \bibinfo {author} {\bibfnamefont {G.}~\bibnamefont {Mileti}},\ }\bibfield  {title} {\enquote {\bibinfo {title} {\ensuremath{\mu}{POP} clock: A microcell atomic clock based on a double-resonance {R}amsey scheme},}\ }\href {https://doi.org/10.1103/PhysRevApplied.18.054039} {\bibfield  {journal} {\bibinfo  {journal} {Phys. Rev. Appl.}\ }\textbf {\bibinfo {volume} {18}},\ \bibinfo {pages} {054039} (\bibinfo {year} {2022})}\BibitemShut {NoStop}%
\bibitem [{\citenamefont {Straessle}\ \emph {et~al.}(2014)\citenamefont {Straessle}, \citenamefont {Pellaton}, \citenamefont {Affolderbach}, \citenamefont {P\'etremand}, \citenamefont {Briand}, \citenamefont {Mileti},\ and\ \citenamefont {de~Rooij}}]{Straessle}%
  \BibitemOpen
  \bibfield  {author} {\bibinfo {author} {\bibfnamefont {R.}~\bibnamefont {Straessle}}, \bibinfo {author} {\bibfnamefont {M.}~\bibnamefont {Pellaton}}, \bibinfo {author} {\bibfnamefont {C.}~\bibnamefont {Affolderbach}}, \bibinfo {author} {\bibfnamefont {Y.}~\bibnamefont {P\'etremand}}, \bibinfo {author} {\bibfnamefont {D.}~\bibnamefont {Briand}}, \bibinfo {author} {\bibfnamefont {G.}~\bibnamefont {Mileti}},\ and\ \bibinfo {author} {\bibfnamefont {N.~F.}\ \bibnamefont {de~Rooij}},\ }\bibfield  {title} {\enquote {\bibinfo {title} {Microfabricated alkali vapor cell with anti-relaxation wall coating},}\ }\href {https://doi.org/10.1063/1.4891248} {\bibfield  {journal} {\bibinfo  {journal} {Applied Physics Letters}\ }\textbf {\bibinfo {volume} {105}} (\bibinfo {year} {2014}),\ 10.1063/1.4891248}\BibitemShut {NoStop}%
\bibitem [{\citenamefont {Giridhar}\ \emph {et~al.}(2023)\citenamefont {Giridhar}, \citenamefont {Karanth}, \citenamefont {Akshaya}, \citenamefont {Elumalai},\ and\ \citenamefont {Sriram}}]{IJP2023}%
  \BibitemOpen
  \bibfield  {author} {\bibinfo {author} {\bibfnamefont {M.~S.}\ \bibnamefont {Giridhar}}, \bibinfo {author} {\bibfnamefont {S.~P.}\ \bibnamefont {Karanth}}, \bibinfo {author} {\bibnamefont {Akshaya}}, \bibinfo {author} {\bibfnamefont {S.}~\bibnamefont {Elumalai}},\ and\ \bibinfo {author} {\bibfnamefont {K.~V.}\ \bibnamefont {Sriram}},\ }\bibfield  {title} {\enquote {\bibinfo {title} {Absorption spectroscopic studies of chip-scale rubidium atomic vapour cells in a compact {3D} printed magneto-optic package},}\ }\href {https://doi.org/10.56042/ijpap.v61i5.71432} {\bibfield  {journal} {\bibinfo  {journal} {Indian Journal of Pure $\&$ Applied Physics}\ } (\bibinfo {year} {2023}),\ 10.56042/ijpap.v61i5.71432}\BibitemShut {NoStop}%
\bibitem [{\citenamefont {Zipfel}\ \emph {et~al.}(2024)\citenamefont {Zipfel}, \citenamefont {Bevington}, \citenamefont {Wright}, \citenamefont {Chalupczak}, \citenamefont {Quick}, \citenamefont {Steele}, \citenamefont {Nicholson},\ and\ \citenamefont {Guarrera}}]{Zipfel2024arxiv}%
  \BibitemOpen
  \bibfield  {author} {\bibinfo {author} {\bibfnamefont {J.~D.}\ \bibnamefont {Zipfel}}, \bibinfo {author} {\bibfnamefont {P.}~\bibnamefont {Bevington}}, \bibinfo {author} {\bibfnamefont {L.}~\bibnamefont {Wright}}, \bibinfo {author} {\bibfnamefont {W.}~\bibnamefont {Chalupczak}}, \bibinfo {author} {\bibfnamefont {G.}~\bibnamefont {Quick}}, \bibinfo {author} {\bibfnamefont {B.}~\bibnamefont {Steele}}, \bibinfo {author} {\bibfnamefont {J.}~\bibnamefont {Nicholson}},\ and\ \bibinfo {author} {\bibfnamefont {V.}~\bibnamefont {Guarrera}},\ }\href@noop {} {\enquote {\bibinfo {title} {Indirect pumping of alkali-metal gases in a miniature silicon-wafer cell},}\ } (\bibinfo {year} {2024}),\ \Eprint {https://arxiv.org/abs/2402.16695} {arXiv:2402.16695 [quant-ph]} \BibitemShut {NoStop}%
\bibitem [{\citenamefont {Zhou}\ \emph {et~al.}(2024)\citenamefont {Zhou}, \citenamefont {Wang}, \citenamefont {Hu}, \citenamefont {Hu}, \citenamefont {Wang}, \citenamefont {Liang}, \citenamefont {Hu}, \citenamefont {Liu},\ and\ \citenamefont {Ye}}]{Zhou2024Measurement}%
  \BibitemOpen
  \bibfield  {author} {\bibinfo {author} {\bibfnamefont {P.}~\bibnamefont {Zhou}}, \bibinfo {author} {\bibfnamefont {Y.}~\bibnamefont {Wang}}, \bibinfo {author} {\bibfnamefont {Z.}~\bibnamefont {Hu}}, \bibinfo {author} {\bibfnamefont {G.}~\bibnamefont {Hu}}, \bibinfo {author} {\bibfnamefont {A.}~\bibnamefont {Wang}}, \bibinfo {author} {\bibfnamefont {Z.}~\bibnamefont {Liang}}, \bibinfo {author} {\bibfnamefont {J.}~\bibnamefont {Hu}}, \bibinfo {author} {\bibfnamefont {L.}~\bibnamefont {Liu}},\ and\ \bibinfo {author} {\bibfnamefont {M.}~\bibnamefont {Ye}},\ }\bibfield  {title} {\enquote {\bibinfo {title} {On-chip integrated non-magnetic heating devices for quantum sensing applications},}\ }\href {https://doi.org/10.1016/j.measurement.2024.114578} {\bibfield  {journal} {\bibinfo  {journal} {Measurement}\ }\textbf {\bibinfo {volume} {232}},\ \bibinfo {pages} {114578} (\bibinfo {year} {2024})}\BibitemShut {NoStop}%
\bibitem [{\citenamefont {Overstolz}\ \emph {et~al.}(2014)\citenamefont {Overstolz}, \citenamefont {Haesler}, \citenamefont {Bergonzi}, \citenamefont {Pezous}, \citenamefont {Clerc}, \citenamefont {Ischer}, \citenamefont {Kaufmann},\ and\ \citenamefont {Despont}}]{Overstolz2014}%
  \BibitemOpen
  \bibfield  {author} {\bibinfo {author} {\bibfnamefont {T.}~\bibnamefont {Overstolz}}, \bibinfo {author} {\bibfnamefont {J.}~\bibnamefont {Haesler}}, \bibinfo {author} {\bibfnamefont {G.}~\bibnamefont {Bergonzi}}, \bibinfo {author} {\bibfnamefont {A.}~\bibnamefont {Pezous}}, \bibinfo {author} {\bibfnamefont {P.-A.}\ \bibnamefont {Clerc}}, \bibinfo {author} {\bibfnamefont {S.}~\bibnamefont {Ischer}}, \bibinfo {author} {\bibfnamefont {J.}~\bibnamefont {Kaufmann}},\ and\ \bibinfo {author} {\bibfnamefont {M.}~\bibnamefont {Despont}},\ }\bibfield  {title} {\enquote {\bibinfo {title} {Wafer scale fabrication of highly integrated rubidium vapor cells},}\ }in\ \href {https://doi.org/10.1109/MEMSYS.2014.6765700} {\emph {\bibinfo {booktitle} {27th International Conference on Micro Electro Mechanical Systems (MEMS)}}}\ (\bibinfo  {publisher} {IEEE},\ \bibinfo {year} {2014})\ pp.\ \bibinfo {pages} {552--555}\BibitemShut {NoStop}%
\bibitem [{\citenamefont {Karlen}\ \emph {et~al.}(2017{\natexlab{a}})\citenamefont {Karlen}, \citenamefont {Gobet}, \citenamefont {Overstolz}, \citenamefont {Haesler},\ and\ \citenamefont {Lecomte}}]{karlen2017lifetime}%
  \BibitemOpen
  \bibfield  {author} {\bibinfo {author} {\bibfnamefont {S.}~\bibnamefont {Karlen}}, \bibinfo {author} {\bibfnamefont {J.}~\bibnamefont {Gobet}}, \bibinfo {author} {\bibfnamefont {T.}~\bibnamefont {Overstolz}}, \bibinfo {author} {\bibfnamefont {J.}~\bibnamefont {Haesler}},\ and\ \bibinfo {author} {\bibfnamefont {S.}~\bibnamefont {Lecomte}},\ }\bibfield  {title} {\enquote {\bibinfo {title} {Lifetime assessment of {RbN}$_3$-filled {MEMS} atomic vapor cells with {A}l$_2${O}$_3$ coating},}\ }\href {https://doi.org/10.1364/OE.25.002187} {\bibfield  {journal} {\bibinfo  {journal} {Optics express}\ }\textbf {\bibinfo {volume} {25}},\ \bibinfo {pages} {2187--2194} (\bibinfo {year} {2017}{\natexlab{a}})}\BibitemShut {NoStop}%
\bibitem [{\citenamefont {Woetzel}\ \emph {et~al.}(2013)\citenamefont {Woetzel}, \citenamefont {Talkenberg}, \citenamefont {Scholtes}, \citenamefont {IJsselsteijn}, \citenamefont {Schultze},\ and\ \citenamefont {Meyer}}]{Woetzel2013SurfaceCoatTech}%
  \BibitemOpen
  \bibfield  {author} {\bibinfo {author} {\bibfnamefont {S.}~\bibnamefont {Woetzel}}, \bibinfo {author} {\bibfnamefont {F.}~\bibnamefont {Talkenberg}}, \bibinfo {author} {\bibfnamefont {T.}~\bibnamefont {Scholtes}}, \bibinfo {author} {\bibfnamefont {R.}~\bibnamefont {IJsselsteijn}}, \bibinfo {author} {\bibfnamefont {V.}~\bibnamefont {Schultze}},\ and\ \bibinfo {author} {\bibfnamefont {H.-G.}\ \bibnamefont {Meyer}},\ }\bibfield  {title} {\enquote {\bibinfo {title} {Lifetime improvement of micro-fabricated alkali vapor cells by atomic layer deposited wall coatings},}\ }\href {https://doi.org/10.1016/j.surfcoat.2013.01.044} {\bibfield  {journal} {\bibinfo  {journal} {Surface and Coatings Technology}\ }\textbf {\bibinfo {volume} {221}},\ \bibinfo {pages} {158–162} (\bibinfo {year} {2013})}\BibitemShut {NoStop}%
\bibitem [{\citenamefont {Karlen}\ \emph {et~al.}(2018{\natexlab{a}})\citenamefont {Karlen}, \citenamefont {Overstolz}, \citenamefont {Gobet}, \citenamefont {Haesler}, \citenamefont {Droz},\ and\ \citenamefont {Lecomte}}]{Karlen2018}%
  \BibitemOpen
  \bibfield  {author} {\bibinfo {author} {\bibfnamefont {S.}~\bibnamefont {Karlen}}, \bibinfo {author} {\bibfnamefont {T.}~\bibnamefont {Overstolz}}, \bibinfo {author} {\bibfnamefont {J.}~\bibnamefont {Gobet}}, \bibinfo {author} {\bibfnamefont {J.}~\bibnamefont {Haesler}}, \bibinfo {author} {\bibfnamefont {F.}~\bibnamefont {Droz}},\ and\ \bibinfo {author} {\bibfnamefont {S.}~\bibnamefont {Lecomte}},\ }\bibfield  {title} {\enquote {\bibinfo {title} {Gold microdiscs as alkali preferential condensation spots for cell clock long-term frequency improvement},}\ }in\ \href@noop {} {\emph {\bibinfo {booktitle} {2018 European Frequency and Time Forum (EFTF)}}}\ (\bibinfo {year} {2018})\ pp.\ \bibinfo {pages} {1--3}\BibitemShut {NoStop}%
\bibitem [{\citenamefont {Griffith}, \citenamefont {Knappe},\ and\ \citenamefont {Kitching}(2010)}]{GriffithOE2010}%
  \BibitemOpen
  \bibfield  {author} {\bibinfo {author} {\bibfnamefont {W.~C.}\ \bibnamefont {Griffith}}, \bibinfo {author} {\bibfnamefont {S.}~\bibnamefont {Knappe}},\ and\ \bibinfo {author} {\bibfnamefont {J.}~\bibnamefont {Kitching}},\ }\bibfield  {title} {\enquote {\bibinfo {title} {Femtotesla atomic magnetometry in a microfabricated vapor cell},}\ }\href {https://doi.org/10.1364/OE.18.027167} {\bibfield  {journal} {\bibinfo  {journal} {Optics Express}\ }\textbf {\bibinfo {volume} {18}},\ \bibinfo {pages} {27167--27172} (\bibinfo {year} {2010})}\BibitemShut {NoStop}%
\bibitem [{\citenamefont {Sebbag}\ \emph {et~al.}(2021)\citenamefont {Sebbag}, \citenamefont {Talker}, \citenamefont {Naiman}, \citenamefont {Barash},\ and\ \citenamefont {Levy}}]{Sebbag2021}%
  \BibitemOpen
  \bibfield  {author} {\bibinfo {author} {\bibfnamefont {Y.}~\bibnamefont {Sebbag}}, \bibinfo {author} {\bibfnamefont {E.}~\bibnamefont {Talker}}, \bibinfo {author} {\bibfnamefont {A.}~\bibnamefont {Naiman}}, \bibinfo {author} {\bibfnamefont {Y.}~\bibnamefont {Barash}},\ and\ \bibinfo {author} {\bibfnamefont {U.}~\bibnamefont {Levy}},\ }\bibfield  {title} {\enquote {\bibinfo {title} {Demonstration of an integrated nanophotonic chip-scale alkali vapor magnetometer using inverse design},}\ }\href {http://dx.doi.org/10.1038/s41377-021-00499-5} {\bibfield  {journal} {\bibinfo  {journal} {Light: Science $\&$amp; Applications}\ }\textbf {\bibinfo {volume} {10}} (\bibinfo {year} {2021})}\BibitemShut {NoStop}%
\bibitem [{\citenamefont {Ma}\ \emph {et~al.}(2022{\natexlab{a}})\citenamefont {Ma}, \citenamefont {Chen}, \citenamefont {Zhao}, \citenamefont {Luo}, \citenamefont {Yu}, \citenamefont {Wang}, \citenamefont {Guo}, \citenamefont {Yang}, \citenamefont {Lin},\ and\ \citenamefont {Jiang}}]{Ma2022}%
  \BibitemOpen
  \bibfield  {author} {\bibinfo {author} {\bibfnamefont {Y.}~\bibnamefont {Ma}}, \bibinfo {author} {\bibfnamefont {Y.}~\bibnamefont {Chen}}, \bibinfo {author} {\bibfnamefont {L.}~\bibnamefont {Zhao}}, \bibinfo {author} {\bibfnamefont {G.}~\bibnamefont {Luo}}, \bibinfo {author} {\bibfnamefont {M.}~\bibnamefont {Yu}}, \bibinfo {author} {\bibfnamefont {Y.}~\bibnamefont {Wang}}, \bibinfo {author} {\bibfnamefont {J.}~\bibnamefont {Guo}}, \bibinfo {author} {\bibfnamefont {P.}~\bibnamefont {Yang}}, \bibinfo {author} {\bibfnamefont {Q.}~\bibnamefont {Lin}},\ and\ \bibinfo {author} {\bibfnamefont {Z.}~\bibnamefont {Jiang}},\ }\bibfield  {title} {\enquote {\bibinfo {title} {The micro-fabrication and performance analysis of non-magnetic heating chip for miniaturized serf atomic magnetometer},}\ }\href {https://doi.org/10.1016/j.jmmm.2022.169495} {\bibfield  {journal} {\bibinfo  {journal} {Journal of Magnetism and Magnetic Materials}\ }\textbf {\bibinfo {volume} {557}},\ \bibinfo {pages} {169495} (\bibinfo {year}
  {2022}{\natexlab{a}})}\BibitemShut {NoStop}%
\bibitem [{\citenamefont {Jiang}\ \emph {et~al.}(2023)\citenamefont {Jiang}, \citenamefont {Zhai}, \citenamefont {Jiang}, \citenamefont {Wang}, \citenamefont {Chen}, \citenamefont {Zhang}, \citenamefont {Wu}, \citenamefont {Zhang}, \citenamefont {Zeng}, \citenamefont {Lin}, \citenamefont {Wang},\ and\ \citenamefont {Jin}}]{Jiang2023APL}%
  \BibitemOpen
  \bibfield  {author} {\bibinfo {author} {\bibfnamefont {M.}~\bibnamefont {Jiang}}, \bibinfo {author} {\bibfnamefont {H.}~\bibnamefont {Zhai}}, \bibinfo {author} {\bibfnamefont {C.}~\bibnamefont {Jiang}}, \bibinfo {author} {\bibfnamefont {J.}~\bibnamefont {Wang}}, \bibinfo {author} {\bibfnamefont {C.}~\bibnamefont {Chen}}, \bibinfo {author} {\bibfnamefont {Q.}~\bibnamefont {Zhang}}, \bibinfo {author} {\bibfnamefont {D.}~\bibnamefont {Wu}}, \bibinfo {author} {\bibfnamefont {B.}~\bibnamefont {Zhang}}, \bibinfo {author} {\bibfnamefont {Z.}~\bibnamefont {Zeng}}, \bibinfo {author} {\bibfnamefont {J.}~\bibnamefont {Lin}}, \bibinfo {author} {\bibfnamefont {Y.}~\bibnamefont {Wang}},\ and\ \bibinfo {author} {\bibfnamefont {P.}~\bibnamefont {Jin}},\ }\bibfield  {title} {\enquote {\bibinfo {title} {Characterization of $^{87}${R}b {MEMS} vapor cells for miniature atomic magnetometers},}\ }\href {https://doi.org/10.1063/5.0149388} {\bibfield  {journal} {\bibinfo  {journal} {Applied Physics Letters}\ }\textbf {\bibinfo
  {volume} {123}} (\bibinfo {year} {2023}),\ 10.1063/5.0149388}\BibitemShut {NoStop}%
\bibitem [{\citenamefont {Tayler}\ \emph {et~al.}(2022)\citenamefont {Tayler}, \citenamefont {Mouloudakis}, \citenamefont {Zetter}, \citenamefont {Hunter}, \citenamefont {Lucivero}, \citenamefont {Bodenstedt}, \citenamefont {Parkkonen},\ and\ \citenamefont {Mitchell}}]{Tayler2022APL}%
  \BibitemOpen
  \bibfield  {author} {\bibinfo {author} {\bibfnamefont {M.~C.~D.}\ \bibnamefont {Tayler}}, \bibinfo {author} {\bibfnamefont {K.}~\bibnamefont {Mouloudakis}}, \bibinfo {author} {\bibfnamefont {R.}~\bibnamefont {Zetter}}, \bibinfo {author} {\bibfnamefont {D.}~\bibnamefont {Hunter}}, \bibinfo {author} {\bibfnamefont {V.~G.}\ \bibnamefont {Lucivero}}, \bibinfo {author} {\bibfnamefont {S.}~\bibnamefont {Bodenstedt}}, \bibinfo {author} {\bibfnamefont {L.}~\bibnamefont {Parkkonen}},\ and\ \bibinfo {author} {\bibfnamefont {M.~W.}\ \bibnamefont {Mitchell}},\ }\bibfield  {title} {\enquote {\bibinfo {title} {Miniature biplanar coils for alkali-metal-vapor magnetometry},}\ }\href {https://doi.org/10.1103/physrevapplied.18.014036} {\bibfield  {journal} {\bibinfo  {journal} {Physical Review Applied}\ }\textbf {\bibinfo {volume} {18}} (\bibinfo {year} {2022}),\ 10.1103/physrevapplied.18.014036}\BibitemShut {NoStop}%
\bibitem [{\citenamefont {Iivanainen}\ \emph {et~al.}(2021)\citenamefont {Iivanainen}, \citenamefont {M\"{a}kinen}, \citenamefont {Zetter}, \citenamefont {Zevenhoven}, \citenamefont {Ilmoniemi},\ and\ \citenamefont {Parkkonen}}]{Iivanainen2021JAP}%
  \BibitemOpen
  \bibfield  {author} {\bibinfo {author} {\bibfnamefont {J.}~\bibnamefont {Iivanainen}}, \bibinfo {author} {\bibfnamefont {A.~J.}\ \bibnamefont {M\"{a}kinen}}, \bibinfo {author} {\bibfnamefont {R.}~\bibnamefont {Zetter}}, \bibinfo {author} {\bibfnamefont {K.~C.~J.}\ \bibnamefont {Zevenhoven}}, \bibinfo {author} {\bibfnamefont {R.~J.}\ \bibnamefont {Ilmoniemi}},\ and\ \bibinfo {author} {\bibfnamefont {L.}~\bibnamefont {Parkkonen}},\ }\bibfield  {title} {\enquote {\bibinfo {title} {A general method for computing thermal magnetic noise arising from thin conducting objects},}\ }\href {https://doi.org/10.1063/5.0050371} {\bibfield  {journal} {\bibinfo  {journal} {Journal of Applied Physics}\ }\textbf {\bibinfo {volume} {130}} (\bibinfo {year} {2021}),\ 10.1063/5.0050371}\BibitemShut {NoStop}%
\bibitem [{\citenamefont {Ma}\ \emph {et~al.}(2022{\natexlab{b}})\citenamefont {Ma}, \citenamefont {Chen}, \citenamefont {Zhao}, \citenamefont {Yu}, \citenamefont {Wang}, \citenamefont {Guo}, \citenamefont {Yang}, \citenamefont {Lin},\ and\ \citenamefont {Jiang}}]{Yintao_Ma_2022}%
  \BibitemOpen
  \bibfield  {author} {\bibinfo {author} {\bibfnamefont {Y.}~\bibnamefont {Ma}}, \bibinfo {author} {\bibfnamefont {Y.}~\bibnamefont {Chen}}, \bibinfo {author} {\bibfnamefont {L.}~\bibnamefont {Zhao}}, \bibinfo {author} {\bibfnamefont {M.}~\bibnamefont {Yu}}, \bibinfo {author} {\bibfnamefont {Y.}~\bibnamefont {Wang}}, \bibinfo {author} {\bibfnamefont {J.}~\bibnamefont {Guo}}, \bibinfo {author} {\bibfnamefont {P.}~\bibnamefont {Yang}}, \bibinfo {author} {\bibfnamefont {Q.}~\bibnamefont {Lin}},\ and\ \bibinfo {author} {\bibfnamefont {Z.}~\bibnamefont {Jiang}},\ }\bibfield  {title} {\enquote {\bibinfo {title} {Accurate determination of alkali atom density based on zero-field magnetic resonance in a single-beam spin-exchange relaxation-free atomic magnetometer},}\ }\href {https://doi.org/10.1088/1361-6501/ac72f9} {\bibfield  {journal} {\bibinfo  {journal} {Measurement Science and Technology}\ }\textbf {\bibinfo {volume} {33}},\ \bibinfo {pages} {105003} (\bibinfo {year} {2022}{\natexlab{b}})}\BibitemShut {NoStop}%
\bibitem [{\citenamefont {Karlen}\ \emph {et~al.}(2018{\natexlab{b}})\citenamefont {Karlen}, \citenamefont {Overstolz}, \citenamefont {Gobet}, \citenamefont {Haesler}, \citenamefont {Droz},\ and\ \citenamefont {Lecomte}}]{Karlen8409005}%
  \BibitemOpen
  \bibfield  {author} {\bibinfo {author} {\bibfnamefont {S.}~\bibnamefont {Karlen}}, \bibinfo {author} {\bibfnamefont {T.}~\bibnamefont {Overstolz}}, \bibinfo {author} {\bibfnamefont {J.}~\bibnamefont {Gobet}}, \bibinfo {author} {\bibfnamefont {J.}~\bibnamefont {Haesler}}, \bibinfo {author} {\bibfnamefont {F.}~\bibnamefont {Droz}},\ and\ \bibinfo {author} {\bibfnamefont {S.}~\bibnamefont {Lecomte}},\ }\bibfield  {title} {\enquote {\bibinfo {title} {Gold microdiscs as alkali preferential condensation spots for cell clock long-term frequency improvement},}\ }in\ \href {https://doi.org/10.1109/EFTF.2018.8409005} {\emph {\bibinfo {booktitle} {2018 European Frequency and Time Forum (EFTF)}}}\ (\bibinfo {year} {2018})\ pp.\ \bibinfo {pages} {91--93}\BibitemShut {NoStop}%
\bibitem [{\citenamefont {des Poi~Mesures}(1969)}]{Mesures1969}%
  \BibitemOpen
  \bibfield  {author} {\bibinfo {author} {\bibfnamefont {C.~I.}\ \bibnamefont {des Poi~Mesures}},\ }\bibfield  {title} {\enquote {\bibinfo {title} {The international practical temperature scale of 1968},}\ }\href {https://doi.org/10.1088/0026-1394/5/2/001} {\bibfield  {journal} {\bibinfo  {journal} {Metrologia}\ }\textbf {\bibinfo {volume} {5}},\ \bibinfo {pages} {35--44} (\bibinfo {year} {1969})}\BibitemShut {NoStop}%
\bibitem [{\citenamefont {{Vasilakis}}(2011)}]{2011PhDT........41V}%
  \BibitemOpen
  \bibfield  {author} {\bibinfo {author} {\bibfnamefont {G.}~\bibnamefont {{Vasilakis}}},\ }\emph {\bibinfo {title} {{Precision measurements of spin interactions with high density atomic vapors}}},\ \href@noop {} {Ph.D. thesis},\ \bibinfo  {school} {Princeton University, New Jersey} (\bibinfo {year} {2011})\BibitemShut {NoStop}%
\bibitem [{\citenamefont {Shah}\ \emph {et~al.}(2007)\citenamefont {Shah}, \citenamefont {Knappe}, \citenamefont {Schwindt},\ and\ \citenamefont {Kitching}}]{shah2007subpicotesla}%
  \BibitemOpen
  \bibfield  {author} {\bibinfo {author} {\bibfnamefont {V.}~\bibnamefont {Shah}}, \bibinfo {author} {\bibfnamefont {S.}~\bibnamefont {Knappe}}, \bibinfo {author} {\bibfnamefont {P.~D.}\ \bibnamefont {Schwindt}},\ and\ \bibinfo {author} {\bibfnamefont {J.}~\bibnamefont {Kitching}},\ }\bibfield  {title} {\enquote {\bibinfo {title} {Subpicotesla atomic magnetometry with a microfabricated vapour cell},}\ }\href@noop {} {\bibfield  {journal} {\bibinfo  {journal} {Nature Photonics}\ }\textbf {\bibinfo {volume} {1}},\ \bibinfo {pages} {649--652} (\bibinfo {year} {2007})}\BibitemShut {NoStop}%
\bibitem [{qus()}]{quspinQZFMGen3}%
  \BibitemOpen
  \href@noop {} {\enquote {\bibinfo {title} {{Q}{Z}{F}{M} {G}en-3; {Q}u{S}pin --- quspin.com},}\ }\bibinfo {howpublished} {\url{https://quspin.com/products-qzfm/}},\ \bibinfo {note} {[Accessed 11-May-2024]}\BibitemShut {NoStop}%
\bibitem [{\citenamefont {C.~B.~Alcock}\ and\ \citenamefont {Horrigan}(1984)}]{AlcockCMQ1984}%
  \BibitemOpen
  \bibfield  {author} {\bibinfo {author} {\bibfnamefont {V.~P.~I.}\ \bibnamefont {C.~B.~Alcock}}\ and\ \bibinfo {author} {\bibfnamefont {M.~K.}\ \bibnamefont {Horrigan}},\ }\bibfield  {title} {\enquote {\bibinfo {title} {Vapour pressure equations for the metallic elements: 298--2500k},}\ }\href {https://doi.org/10.1179/cmq.1984.23.3.309} {\bibfield  {journal} {\bibinfo  {journal} {Canadian Metallurgical Quarterly}\ }\textbf {\bibinfo {volume} {23}},\ \bibinfo {pages} {309--313} (\bibinfo {year} {1984})},\ \Eprint {https://arxiv.org/abs/https://doi.org/10.1179/cmq.1984.23.3.309} {https://doi.org/10.1179/cmq.1984.23.3.309} \BibitemShut {NoStop}%
\bibitem [{\citenamefont {Haynes}(2016)}]{CRCHandbook97thEdition}%
  \BibitemOpen
  \bibinfo {editor} {\bibfnamefont {W.~M.}\ \bibnamefont {Haynes}},\ ed.,\ \href@noop {} {\emph {\bibinfo {title} {CRC Handbook of Chemistry and Physics}}},\ \bibinfo {edition} {97th}\ ed.\ (\bibinfo  {publisher} {Taylor \& Francis},\ \bibinfo {year} {2016})\ \bibinfo {note} {iSBN 9781315380476}\BibitemShut {NoStop}%
\bibitem [{\citenamefont {Killian}(1926)}]{KillianPR1926}%
  \BibitemOpen
  \bibfield  {author} {\bibinfo {author} {\bibfnamefont {T.~J.}\ \bibnamefont {Killian}},\ }\bibfield  {title} {\enquote {\bibinfo {title} {Thermionic phenomena caused by vapors of rubidium and potassium},}\ }\href {https://doi.org/10.1103/PhysRev.27.578} {\bibfield  {journal} {\bibinfo  {journal} {Phys. Rev.}\ }\textbf {\bibinfo {volume} {27}},\ \bibinfo {pages} {578--587} (\bibinfo {year} {1926})}\BibitemShut {NoStop}%
\bibitem [{\citenamefont {Seltzer}(2008)}]{SeltzerThesis}%
  \BibitemOpen
  \bibfield  {author} {\bibinfo {author} {\bibfnamefont {S.~J.}\ \bibnamefont {Seltzer}},\ }\emph {\bibinfo {title} {Developments in Alkali-Metal Atomic Magnetometry}},\ \href {http://physics.princeton.edu/romalis/papers/Seltzer%20Thesis.pdf} {Ph.D. thesis},\ \bibinfo  {school} {Princeton University} (\bibinfo {year} {2008})\BibitemShut {NoStop}%
\bibitem [{\citenamefont {Singh}, \citenamefont {Dolph},\ and\ \citenamefont {Tobias}(2008)}]{SinghPDF2008}%
  \BibitemOpen
  \bibfield  {author} {\bibinfo {author} {\bibfnamefont {J.}~\bibnamefont {Singh}}, \bibinfo {author} {\bibfnamefont {P.~A.~M.}\ \bibnamefont {Dolph}},\ and\ \bibinfo {author} {\bibfnamefont {W.~A.}\ \bibnamefont {Tobias}},\ }\href {https://people.nscl.msu.edu/~singhj/docs/vp195.pdf} {\enquote {\bibinfo {title} {Alkali metal vapor pressures i\& number densities for hybrid spin exchange optical pumping},}\ } (\bibinfo {year} {2008})\BibitemShut {NoStop}%
\bibitem [{\citenamefont {Romalis}, \citenamefont {Miron},\ and\ \citenamefont {Cates}(1997)}]{PhysRevA.56.4569}%
  \BibitemOpen
  \bibfield  {author} {\bibinfo {author} {\bibfnamefont {M.~V.}\ \bibnamefont {Romalis}}, \bibinfo {author} {\bibfnamefont {E.}~\bibnamefont {Miron}},\ and\ \bibinfo {author} {\bibfnamefont {G.~D.}\ \bibnamefont {Cates}},\ }\bibfield  {title} {\enquote {\bibinfo {title} {Pressure broadening of {R}b {D}$_1$ and {D}$_2$ lines by ${}^{3}${H}e, ${}^{4}${H}e, {N}${}_{2}$, and {Xe}: Line cores and near wings},}\ }\href {https://doi.org/10.1103/PhysRevA.56.4569} {\bibfield  {journal} {\bibinfo  {journal} {Phys. Rev. A}\ }\textbf {\bibinfo {volume} {56}},\ \bibinfo {pages} {4569--4578} (\bibinfo {year} {1997})}\BibitemShut {NoStop}%
\bibitem [{\citenamefont {Krzyzewski}\ \emph {et~al.}(2019)\citenamefont {Krzyzewski}, \citenamefont {Perry}, \citenamefont {Gerginov},\ and\ \citenamefont {Knappe}}]{Krzyzewski2019}%
  \BibitemOpen
  \bibfield  {author} {\bibinfo {author} {\bibfnamefont {S.~P.}\ \bibnamefont {Krzyzewski}}, \bibinfo {author} {\bibfnamefont {A.~R.}\ \bibnamefont {Perry}}, \bibinfo {author} {\bibfnamefont {V.}~\bibnamefont {Gerginov}},\ and\ \bibinfo {author} {\bibfnamefont {S.}~\bibnamefont {Knappe}},\ }\bibfield  {title} {\enquote {\bibinfo {title} {Characterization of noise sources in a microfabricated single-beam zero-field optically-pumped magnetometer},}\ }\href {https://doi.org/10.1063/1.5098088} {\bibfield  {journal} {\bibinfo  {journal} {Journal of Applied Physics}\ }\textbf {\bibinfo {volume} {126}} (\bibinfo {year} {2019}),\ 10.1063/1.5098088}\BibitemShut {NoStop}%
\bibitem [{\citenamefont {Lucivero}\ \emph {et~al.}(2022)\citenamefont {Lucivero}, \citenamefont {Zanoni}, \citenamefont {Corrielli}, \citenamefont {Osellame},\ and\ \citenamefont {Mitchell}}]{LuciveroLWVC}%
  \BibitemOpen
  \bibfield  {author} {\bibinfo {author} {\bibfnamefont {V.~G.}\ \bibnamefont {Lucivero}}, \bibinfo {author} {\bibfnamefont {A.}~\bibnamefont {Zanoni}}, \bibinfo {author} {\bibfnamefont {G.}~\bibnamefont {Corrielli}}, \bibinfo {author} {\bibfnamefont {R.}~\bibnamefont {Osellame}},\ and\ \bibinfo {author} {\bibfnamefont {M.~W.}\ \bibnamefont {Mitchell}},\ }\bibfield  {title} {\enquote {\bibinfo {title} {Laser-written vapor cells for chip-scale atomic sensing and spectroscopy},}\ }\href {https://doi.org/10.1364/OE.469296} {\bibfield  {journal} {\bibinfo  {journal} {Opt. Express}\ }\textbf {\bibinfo {volume} {30}},\ \bibinfo {pages} {27149--27163} (\bibinfo {year} {2022})}\BibitemShut {NoStop}%
\bibitem [{\citenamefont {Zanoni}\ \emph {et~al.}(2024)\citenamefont {Zanoni}, \citenamefont {Mouloudakis}, \citenamefont {Tayler}, \citenamefont {Corrielli}, \citenamefont {Osellame}, \citenamefont {Mitchell},\ and\ \citenamefont {Lucivero}}]{zanoni2024laserwritten}%
  \BibitemOpen
  \bibfield  {author} {\bibinfo {author} {\bibfnamefont {A.}~\bibnamefont {Zanoni}}, \bibinfo {author} {\bibfnamefont {K.}~\bibnamefont {Mouloudakis}}, \bibinfo {author} {\bibfnamefont {M.~C.~D.}\ \bibnamefont {Tayler}}, \bibinfo {author} {\bibfnamefont {G.}~\bibnamefont {Corrielli}}, \bibinfo {author} {\bibfnamefont {R.}~\bibnamefont {Osellame}}, \bibinfo {author} {\bibfnamefont {M.~W.}\ \bibnamefont {Mitchell}},\ and\ \bibinfo {author} {\bibfnamefont {V.~G.}\ \bibnamefont {Lucivero}},\ }\href@noop {} {\enquote {\bibinfo {title} {Laser-written micro-channel atomic magnetometer},}\ } (\bibinfo {year} {2024}),\ \Eprint {https://arxiv.org/abs/2404.14345} {arXiv:2404.14345 [quant-ph]} \BibitemShut {NoStop}%
\bibitem [{\citenamefont {Mouloudakis}\ \emph {et~al.}(2024)\citenamefont {Mouloudakis}, \citenamefont {Koutrouli}, \citenamefont {Kominis}, \citenamefont {Mitchell},\ and\ \citenamefont {Vasilakis}}]{mouloudakis2024spin}%
  \BibitemOpen
  \bibfield  {author} {\bibinfo {author} {\bibfnamefont {K.}~\bibnamefont {Mouloudakis}}, \bibinfo {author} {\bibfnamefont {V.}~\bibnamefont {Koutrouli}}, \bibinfo {author} {\bibfnamefont {I.~K.}\ \bibnamefont {Kominis}}, \bibinfo {author} {\bibfnamefont {M.~W.}\ \bibnamefont {Mitchell}},\ and\ \bibinfo {author} {\bibfnamefont {G.}~\bibnamefont {Vasilakis}},\ }\href@noop {} {\enquote {\bibinfo {title} {Spin projection noise and the magnetic sensitivity of optically pumped magnetometers},}\ } (\bibinfo {year} {2024}),\ \Eprint {https://arxiv.org/abs/2402.10746} {arXiv:2402.10746 [quant-ph]} \BibitemShut {NoStop}%
\bibitem [{\citenamefont {Karlen}\ \emph {et~al.}(2017{\natexlab{b}})\citenamefont {Karlen}, \citenamefont {Gobet}, \citenamefont {Overstolz}, \citenamefont {Haesler},\ and\ \citenamefont {Lecomte}}]{Karlen2017AppliedSpec}%
  \BibitemOpen
  \bibfield  {author} {\bibinfo {author} {\bibfnamefont {S.}~\bibnamefont {Karlen}}, \bibinfo {author} {\bibfnamefont {J.}~\bibnamefont {Gobet}}, \bibinfo {author} {\bibfnamefont {T.}~\bibnamefont {Overstolz}}, \bibinfo {author} {\bibfnamefont {J.}~\bibnamefont {Haesler}},\ and\ \bibinfo {author} {\bibfnamefont {S.}~\bibnamefont {Lecomte}},\ }\bibfield  {title} {\enquote {\bibinfo {title} {Quantitative micro-{R}aman spectroscopy for partial pressure measurement in small volumes},}\ }\href {https://doi.org/10.1177/0003702817724410} {\bibfield  {journal} {\bibinfo  {journal} {Applied Spectroscopy}\ }\textbf {\bibinfo {volume} {71}},\ \bibinfo {pages} {2707–2713} (\bibinfo {year} {2017}{\natexlab{b}})}\BibitemShut {NoStop}%
\bibitem [{\citenamefont {Wu}\ \emph {et~al.}(1986)\citenamefont {Wu}, \citenamefont {Kitano}, \citenamefont {Happer}, \citenamefont {Hou},\ and\ \citenamefont {Daniels}}]{WuAO1986}%
  \BibitemOpen
  \bibfield  {author} {\bibinfo {author} {\bibfnamefont {Z.}~\bibnamefont {Wu}}, \bibinfo {author} {\bibfnamefont {M.}~\bibnamefont {Kitano}}, \bibinfo {author} {\bibfnamefont {W.}~\bibnamefont {Happer}}, \bibinfo {author} {\bibfnamefont {M.}~\bibnamefont {Hou}},\ and\ \bibinfo {author} {\bibfnamefont {J.}~\bibnamefont {Daniels}},\ }\bibfield  {title} {\enquote {\bibinfo {title} {Optical determination of alkali metal vapor number density using {F}araday rotation},}\ }\href {https://doi.org/10.1364/AO.25.004483} {\bibfield  {journal} {\bibinfo  {journal} {Appl. Opt.}\ }\textbf {\bibinfo {volume} {25}},\ \bibinfo {pages} {4483--4492} (\bibinfo {year} {1986})}\BibitemShut {NoStop}%
\bibitem [{\citenamefont {Edri}\ \emph {et~al.}(2021)\citenamefont {Edri}, \citenamefont {Armon}, \citenamefont {Greenberg}, \citenamefont {Moshe-Tsurel}, \citenamefont {Lubotzky}, \citenamefont {Salzillo}, \citenamefont {Perelshtein}, \citenamefont {Tkachev}, \citenamefont {Girshevitz},\ and\ \citenamefont {Shpaisman}}]{Edri2021ACSappliedmaterials}%
  \BibitemOpen
  \bibfield  {author} {\bibinfo {author} {\bibfnamefont {E.}~\bibnamefont {Edri}}, \bibinfo {author} {\bibfnamefont {N.}~\bibnamefont {Armon}}, \bibinfo {author} {\bibfnamefont {E.}~\bibnamefont {Greenberg}}, \bibinfo {author} {\bibfnamefont {S.}~\bibnamefont {Moshe-Tsurel}}, \bibinfo {author} {\bibfnamefont {D.}~\bibnamefont {Lubotzky}}, \bibinfo {author} {\bibfnamefont {T.}~\bibnamefont {Salzillo}}, \bibinfo {author} {\bibfnamefont {I.}~\bibnamefont {Perelshtein}}, \bibinfo {author} {\bibfnamefont {M.}~\bibnamefont {Tkachev}}, \bibinfo {author} {\bibfnamefont {O.}~\bibnamefont {Girshevitz}},\ and\ \bibinfo {author} {\bibfnamefont {H.}~\bibnamefont {Shpaisman}},\ }\bibfield  {title} {\enquote {\bibinfo {title} {Laser printing of multilayered alternately conducting and insulating microstructures},}\ }\href {https://doi.org/10.1021/acsami.1c06204} {\bibfield  {journal} {\bibinfo  {journal} {ACS Applied Materials $\&$; Interfaces}\ }\textbf {\bibinfo {volume} {13}},\ \bibinfo {pages} {36416–36425} (\bibinfo
  {year} {2021})}\BibitemShut {NoStop}%
\end{thebibliography}%
\end{document}